\documentclass[journal=jacsat,manuscript=article]{achemso}

\usepackage{chemformula} 
\usepackage[T1]{fontenc} 

\author{Qiang Zhang}
\affiliation{Neutron Scattering Division, Oak Ridge National Laboratory, Oak Ridge, Tennessee 37831, USA}
\email{zhangq6@ornl.gov}
\author{Yuanpeng Zhang}
\affiliation{Neutron Scattering Division, Oak Ridge National Laboratory, Oak Ridge, Tennessee 37831, USA}
\email{zhangy3@ornl.gov}

\author{Masaaki Matsuda}
\affiliation{Neutron Scattering Division, Oak Ridge National Laboratory, Oak Ridge, Tennessee 37831, USA}
 
\author{Vasile Ovidiu Garlea}
\affiliation{Neutron Scattering Division, Oak Ridge National Laboratory, Oak Ridge, Tennessee 37831, USA}
\author{Jiaqiang Yan}
\affiliation{Materials Science and Technology Division, Oak Ridge National Laboratory, Oak Ridge, Tennessee 37831, USA}
\author{Michael A. McGuire}
\affiliation{Materials Science and Technology Division, Oak Ridge National Laboratory, Oak Ridge, Tennessee 37831, USA}
 \author{D. Alan Tennant}
\affiliation{Materials Science and Technology Division, Oak Ridge National Laboratory, Oak Ridge, Tennessee 37831, USA}
\alsoaffiliation{Quantum Science Center, Oak Ridge, Tennessee 37831, USA}
\alsoaffiliation{Shull Wollan Center, Oak Ridge National Laboratory, Oak Ridge, Tennessee 37831, USA }
\altaffiliation{Current address: Dept. of Physics and Astronomy, University of Tennessee Knoxville, TN 37996-1200}
\author{Satoshi Okamoto}
\affiliation{Materials Science and Technology Division, Oak Ridge National Laboratory, Oak Ridge, Tennessee 37831, USA}
 
\title[An \textsf{achemso} demo]
  {Hidden local symmetry breaking in a kagome-lattice magnetic Weyl semimetal
  }

\abbreviations{IR,NMR,UV}
\keywords{American Chemical Society, \LaTeX}
 
\begin{document}
\begin{tocentry}

 \begin{center}
     \includegraphics[width=8.5cm]{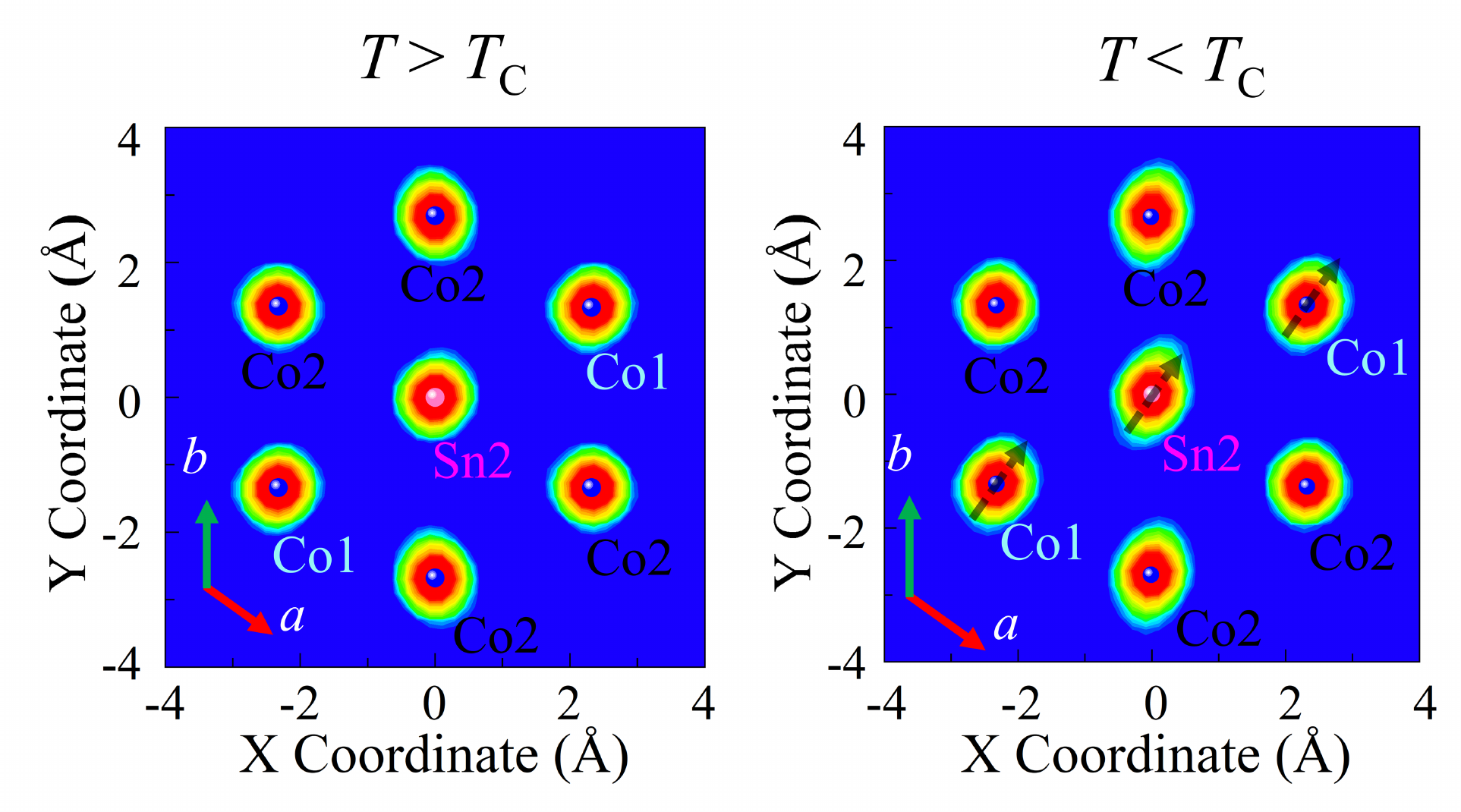}
%
 \end{center}
 
%
 

\end{tocentry}

\begin{abstract}
  Exploring the relationship between intriguing physical properties and structural complexity is a central topic in studying modern functional materials. Co$_{3}$Sn$_{2}$S$_{2}$, a new discovered kagome-lattice magnetic Weyl semimetal, has triggered intense interest owing to the intimate coupling between topological semimetallic states and peculiar magnetic properties. However, the origins of the magnetic phase separation and spin glass state below $T_{C}$ in this ordered compound are two unresolved yet important puzzles in understanding its magnetism. Here, we report the discovery of local symmetry breaking surprisingly co-emerges with the onset of ferromagnetic order in Co$_{3}$Sn$_{2}$S$_{2}$, by a combined use of neutron total scattering and half polarized neutron diffraction. An anisotropic distortion of the cobalt kagome lattice at atomic/nano level is also found, with distinct distortion directions among the two Co1 and four Co2 atoms. The mismatch of local and average symmetries occurs below $T_{C}$, indicating that Co$_{3}$Sn$_{2}$S$_{2}$ evolves to an intrinsically lattice disordered system when the ferromagnetic order is established. The local symmetry breaking with intrinsic lattice disorder provides new understandings to the puzzling magnetic properties. Our density function theory (DFT) calculation indicates that the local symmetry breaking is expected to reorient local ferromagnetic moments, unveiling the existence of the ferromagnetic instability associated with the lattice instability. Furthermore, DFT calculation unveils that the local symmetry breaking could affect the Weyl property by breaking mirror plane. Our findings highlight the fundamentally important role that the local symmetry breaking plays in advancing our understanding on the magnetic and topological properties in Co$_{3}$Sn$_{2}$S$_{2}$, which may draw the attention to explore the overlooked local symmetry breaking in Co$_{3}$Sn$_{2}$S$_{2}$, its derivatives, and more broadly in other topological Dirac/Weyl semimetals and kagome-lattice magnets. 
  
\end{abstract}

\section{Introduction}
 
Local structure deviated from average structure is crucial to understand the macroscopic physical properties in modern functional materials\cite{Egami2012,Keen2015,Zhu2021}. In a doped alloy or even ordered, non-alloyed crystal, a local structure manifested by local atomic displacement or polyhedral distortion may be different from the average crystal structure.\cite{Egami2012,Wang2020,Zhu2021}. Local symmetry breaking was found in some of doped nonmagnetic\cite{Mikkelse1982,EGAMI1991,Chong2012} or magnetic\cite{Bianconi1996,Despina1997,Perversi2019,Frandsen2017,Jiang2021} alloys, the origin of which is related to the different local configurations arising from the site disorder extrinsically introduced through chemical doping or nonstoichiometry.  Relatively few ordered and non-alloyed magnetic materials without chemical doping or nonstoichiometry\cite{Goodwin2006,Allieta2012,Keen2015,Lu2017,Wang2020} also exhibit a distinct lower symmetry from the average structure. The local symmetry breaking in such magnetic materials may be more interesting since it is not associated with the chemical doping/nonstoichiometry and reflects an intimate coupling between lattice and internal forces such as spins or orbitals. Recent investigations on such magnetic materials revealed that the local symmetry breaking is indispensable to provide a valid insight to their physical properties. For example, a canted ferromagnetic order is preceded by a local cubic symmetry breaking at higher temperature in Ba$_{2}$NaOsO$_{6}$, which provides an experimental confirmation of the microscopic quantum models\cite{Lu2017}. Note that a common feature\cite{Allieta2012,Frandsen2017} in the reported magnetic materials is that the local symmetry breaking occurs at a temperature higher than magnetic transition temperature, and the long-range structural transition temperature where the average lattice symmetry usually breaks to be the same to the local symmetry. Furthermore, from theoretical perspective, it may not be a safe practice to proceed the electronic band structure or other theoretical calculations using the phenomenological models that are based on macroscopic (crystallographic) parameters only without considering the distinct local structure.\cite{Wang2020,Trimarchi2018} Unfortunately, current understandings on the two large categories of magnetic Dirac/Weyl semimetals\cite{Joonbum2011, Liu2017, Nakatsuji2015,Nagaosa2020} and kagome-lattice magnets \cite{Lucile2016,Zhang2018} are mainly based on the globally averaged crystal structure\cite{Joonbum2011, Liu2017, Nakatsuji2015,Yan2017,Lucile2016,Zhang2018,Nagaosa2020,Zou2019} and the hidden local symmetry breaking remains elusive in these materials. More importantly, the relationship among local symmetry breaking, average symmetry breaking and magnetic transition, and the impact of the local symmetry breaking on the local/macroscopic magnetism and the topological properties in these materials have not been explored.

The kagome Weyl magnet Co$_{3}$Sn$_{2}$S$_{2}$\cite{Liu2018,Liu2019} has attracted much attention due to the novel topological states\cite{Yin2019, Morali2019,Yang2020} coupled to the peculiar magnetism\cite{Kassem2017,Schnelle2013,Guguchia2020,Lachman2020,Zhang2021}. Despite numerous theoretical and experimental studies of this important compound, the
magnetic phase separation\cite{Guguchia2020}  and spin glass state\cite{Lachman2020} in Co$_{3}$Sn$_{2}$S$_{2}$ are two intriguing but puzzling magnetic properties. Co$_{3}$Sn$_{2}$S$_{2}$ is an ordered compound without detectable chemical doping, nonstoichiometry, or antisite and has been considered to retain rhombohedral structure with space group $R\mbox{-}3m$ down to $\sim$ 0 K. \cite{Liu2018,Liu2019,Yin2019, Morali2019,Yang2020,Kassem2017,Schnelle2013,Guguchia2020,Lachman2020,Zhang2021} The magnetic phase separation and spin glass state in Co$_{3}$Sn$_{2}$S$_{2}$ are two puzzling magnetic properties. Co$_{3}$Sn$_{2}$S$_{2}$ exhibits two anomalies at $T_{A}\approx135$ K and $T_{C}\approx175$ K in the susceptibility\cite{Schnelle2013,Kassem2017}. The ground state of Co$_{3}$Sn$_{2}$S$_{2}$ is consistent with a ferromagnetic order with the easy axis along the out-of-plane $c_{H}$ axis\cite{Schnelle2013,Kassem2017,Zhang2021}. Interestingly, Guguchia et al.\cite{Guguchia2020} reported a robust magnetic phase separation of the out-of-plane FM order and a  120$^{\rm o}$ in-plane antiferromagnetic(AFM) order in $T_{A}<T<T_{C}$, followed by a sole FM order in $T<T_{A}$ based on the local probe technique Muon spin rotation ($\mu SR$) as shown in Fig. 1(a). The FM volume fraction starts to decrease with increasing the temperature above $T_{A}$, with a compensation of the AFM volume fraction.  The proposed $R\mbox{-}3m$ 120$^{\rm o}$ order was not confirmed by the spherical neutron polarimetry analysis and another type of $R\mbox{-}3m'$ 120$^{\rm o}$ order may be likely\cite{Soh2021}. The two separated FM and AFM phases in $T_{A}<T<T_{C}$ are puzzling since they belong to two different irreducible representations \cite{Guguchia2020} and are not allowed by a simple rhombohedral structure with only one Co site from the perspective of the symmetry analysis. Furthermore, a spin glass state\cite{Lachman2020} starts to appear at $T_{C}$ and becomes stiffer below $T_{A}$. A coexistence of the stiffer spin glass state and FM phase was proposed\cite{Lachman2020} to interpret the exchange biased AHE and magnetic hysteresis loops in $T<T_{A}$.  The existence of the spin glass state raised questions if the ferromagnetism is short-range or long-range ordered and whether the ferromagnetic moment direction remains along the $c_{H}$ axis in $T<T_{C}$. Since the ferromagnetic moment is very low ( $<0.35 \mu_{B}$), it is difficult for unpolarized neutron diffraction technique\cite{VAQUEIRO2009513,Zhang2021}. Furthermore, although there exist an intrinsic geometric frustration to the cobalt kagome lattice and a weak spin frustration\cite{Zhang2021}, the appearance of the spin glass state in  $T<T_{C}$ in the well-ordered rhombohedral structure\cite{VAQUEIRO2009513,Zhang2021} is surprising since there lacks the lattice/site disorder that is generally needed for the formation of spin glass state\cite{Mydosh1993}. It is also unknown why the spin glass state becomes stiffer below $T_{A}$ to be responsible for the appearance of exchange bias. To understand the puzzling magnetic phase separation and spin glass state in Co$_{3}$Sn$_{2}$S$_{2}$, a comprehensive understanding on the crystal structure and ferromagnetic order in both macroscopic and microscopic (local) scales are required. Given that the average rhombohedral structure is maintained down to $\approx$ 0 K \cite{Liu2018,Liu2019,Yin2019, Morali2019,Yang2020,Kassem2017,Schnelle2013,Guguchia2020,Lachman2020,Zhang2021} and there are successive magnetic transitions at $T_{C}$ and $T_{A}$, it is of great interest to explore if there is any change in the local symmetry at atomic/nano scale associated with these two magnetic transitions, which may in turn influence the local moment configuration, macroscopic magnetic properties and electronic band topology.

In this paper, we report a local symmetry breaking and its novel correlations to the magnetism and topological properties in Co$_{3}$Sn$_{2}$S$_{2}$. Neutron diffraction shows an anomaly in the in-plane lattice constant and average Co-Sn2-Co bond length of the global rhombohedral structure at $T_{C}$. The advanced reverse Monte Carlo (RMC) approach to model high-quality neutron total scattering results reveals a striking local symmetry breaking from rhombohedral $R\mbox{-}3m$ to monoclinic $Cm$ below ferromagnetic $T_{C}$ with an unchanged average rhombohedral structure down to the lowest investigated temperature 6 K. The phenomena is different from that in other magnetic materials, revealing a novel coupling between lattice and spin degree of freedoms in Co$_{3}$Sn$_{2}$S$_{2}$. An anisotropic cobalt kagome lattice distortion is found below $T_{C}$, in which the in-plane projection of the local atomic displacements for two Co1 atoms points to the +/-(100) direction but those of
four Co2 atom do not point to any high symmetric directions. The local symmetry breaking with the intrinsic lattice disorder provides new understandings to the previously puzzling magnetic phase separation and spin glass like state in Co$_{3}$Sn$_{2}$S$_{2}$. The average long-range FM moment points to the c$_{H}$ axis as confirmed from our half-polarized neutron diffraction experiments. However, the local symmetry breaking leads to a local ferromagnetic instability and has a tendency to drive a local FM moment reorientation by $\approx$ 20$^{\rm o}$ relative to the $c_{H}$ axis. Furthermore, our density function theory calculations reveal that the local monoclinic distortion plays a detrimental role in the formation of the Weyl points by breaking mirror symmetries and could induce broad topological surface band like feature. Our results make Co$_{3}$Sn$_{2}$S$_{2}$ the first example in two large categories of magnetic semimetals and kagome magnets exhibiting a locally lower symmetry than average lattice symmetry that coemerges with the FM order and correlates to the topological properties. 

\begin{figure}
\centering
\includegraphics[width=0.75\linewidth]{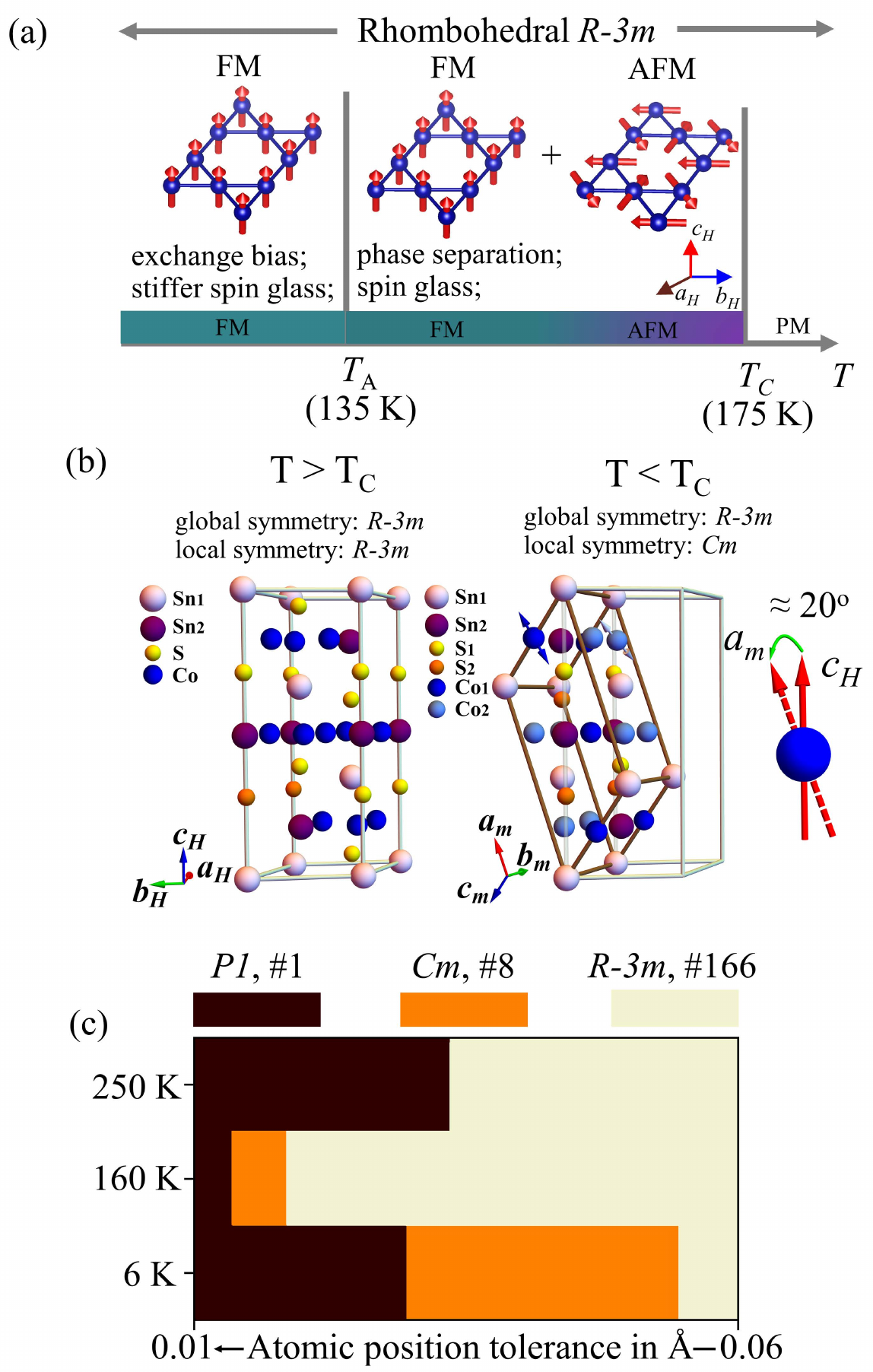}
\caption{(a). Crystal structure and the magnetic properties of Co$_{3}$Sn$_{2}$S$_{2}$ based on previous reports. The color change of the horizontal bar in $T_{A}<T<T_{C}$ indicates the temperature evolution of the FM volume fraction (aqua) and AFM volume fraction (purple). (b). The global and local symmetries of Co$_{3}$Sn$_{2}$S$_{2}$ in $T> T_{C}$ and $T< T_{C}$ temperature regions. The gray hexagonal unit cell is also displayed in the middle panel to show the conversion between the hexagonal $R\mbox{-}3m$ and distorted monoclinic $Cm$ unit cells. The arrows in the monoclinic unit cell illustrate the schematic displacements of Co1 and Co2 atoms in the $a_{m}c_{m}$ plane and one 3D direction, respectively. The right panel illustrates a tendency of the local spin reorientation from the $c_{H}$ axis to the $a_{m}$ axis in $T< T_{C}$. (c). Space group extraction from collapsed unit cells for the neutron diffraction and PDF datasets at 250, 160 and 6 K in Co$_{3}$Sn$_{2}$S$_{2}$. }
\label{fig:1}
\end{figure}

\section{Results}
\subsection{Average ferromagnetic order and moment from half-polarized neutron diffraction}

Half-polarized neutron diffraction was employed to study the ferromagnetic order which has a low magnetic moment in Co$_{3}$Sn$_{2}$S$_{2}$. In contrast to the unpolarized scattering from a weak ferromagnetic component that produces only a minor change in the intensity of nuclear peaks, the half polarized scattering allows to enhance the sensitivity to the magnetic signal by taking advantage of the interference term P$_{0}$*F$_{M}$*F$_{N}$, where F$_{N}$ and F$_{M}$ are the nuclear and magnetic structure factors, and P$_{0}$ is the incident neutron beam polarization. A vertical external magnetic field was applied to co-align the ferromagnetic domains and avoid the beam depolarization. For the setup with $\textbf{Q}\bot c_{H}$ in the hexagonal notation, the up-spin and down-spin diffraction intensities I$^{+}$ and I$^{-}$ for the  Q=(110) peak are described as (F$_{N}^2$+2P$_{0}\times$F$_{N}$F$_{M}$+F$_{M}^2$) and (F$_{N}^2$-2P$_{0}\times$F$_{N}$F$_{M}$+F$_{M}^2$), respectively.\cite{Leli2010} Thus, the flipping ratio R=I$^{+}$/I$^{-}\approx$ 1+4P$_{0}\times$F$_{M}$/F$_{N}$. On the other hand, the difference between I$^{+}$ and I$^{-}$ is 4P$_{0}\times$F$_{N}$F$_{M}$, which is proportional to F$_{M}$. Therefore, even if the FM moment is small, it can be detected by the flipping ratio R or flipping difference (I$^{+}$-I$^{-}$).

\begin{figure}
\centering
\includegraphics[width=1\linewidth]{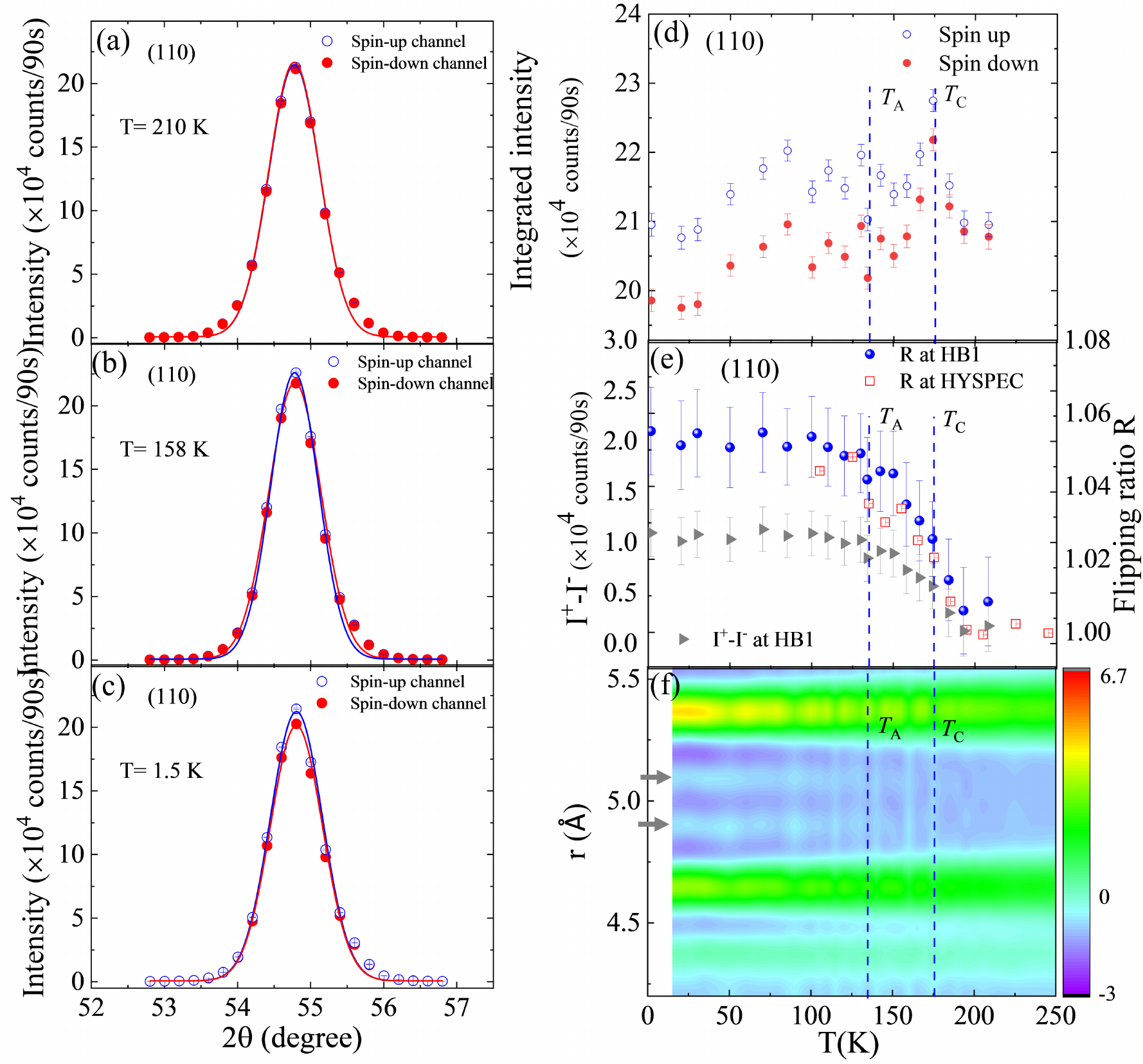}
\caption{Half-polalrized neutron diffraction results from the HB1 triple-axis spectrometer: $\theta-2\theta$ scan of the up-spin diffraction intensity I$^{+}$ and down-spin diffraction intensity I$^{-}$ of (110) peak at (a) 210 K, (b) 158 K and (c) 1.5 K. Temperature dependence of (d) the integrated intensity of I$^{+}$ and I$^{-}$, (e) the flipping ratios R=I$^{+}$/I$^{-}$ (HYSPEC data was added for comparison), and the difference I$^{+}$-I$^{-}$. (f) Contour plot of the PDF patterns vs temperature. The two arrows indicate the locations of the two new PDF peaks appearing below $T_{C}$}
\label{fig:2}
\end{figure}
Figure~2 (a-c) shows the $\theta-2\theta$ scan of the up-spin diffraction intensity I$^{+}$ and down-spin diffraction intensity I$^{-}$ of (110) peak at 210 K ($T>T_{C}$), 158 K ($T_{A}<T<T_{C}$) and 1.5 K ($T<T_{A}$) collected at the triple axis spectrometer HB1, with P$_{0}$ of $\approx$80\%. At 210 K, I$^{+}$ and I$^{-}$ show no difference indicating there is no ferromagnetic signal. At 158 K, there is a difference in I$^{+}$ and I$^{-}$, which becomes larger at 1.5 K.  The temperature dependence of the integrated intensity for  I$^{+}$ and I$^{-}$ is displayed in Fig. 2 (d). The flipping ratio R and difference of I$^{+}$ and I$^{-}$ are shown in Fig. ~2(e). The flipping ratio derived from HYSPEC is over-plotted in Fig. ~2(e), showing a good consistency. Both the flipping ratios and difference of (110) peak exhibit a clear increase below around $T_{C}$ and become unchanged near $T_{A}$. This indicates that (110) is the ferromagnetic peak that appears below $T_{C}$ and becomes saturated below $T_{A}$. There is no broadening in the linewidths of the peaks for spin-up and spin-down channels going through $T_{C}$ and $T_{A}$, indicating that ferromagnetic order is of long range. At base temperature 4 K, the flipping ratio is 1.055. We obtained F$_{M}$/F$_{N}$ $\approx$ (R-1)/(4$\times$P$_{0}$)=0.0172 yielding  F$^2_{M}$/F$^2_{N}$ $\approx$ 0.00030. The ordered moment of the long-range ferromagnetic order is determined to be $\approx$ 0.14(2)$\mu_{B}$ at 1.5 K, with the moment along the $c_{H}$ axis. The long-range ordered FM moment size is consistent with the magnitude based on the microscopic $\mu SR$ measurements\cite{Guguchia2020}, but much lower than the saturation moment $\approx$ 0.35$\mu_{B}$ derived from the magnetization measurements\cite{Schnelle2013,Kassem2017,Zhang2021}. As will be discussed below, the local ferromagnetic instability may reconcile such discrepancy. Thus, in $T<T_{C}$, the average FM order remains long-range ordered along $c_{H}$, which coexists with the previously reported spin glass state\cite{Lachman2020} in Co$_{3}$Sn$_{2}$S$_{2}$.

\subsection{Temperature dependence of lattice constants and average bond length/angles from neutron diffraction}
 \begin{figure}
\centering
\includegraphics[width=1\linewidth]{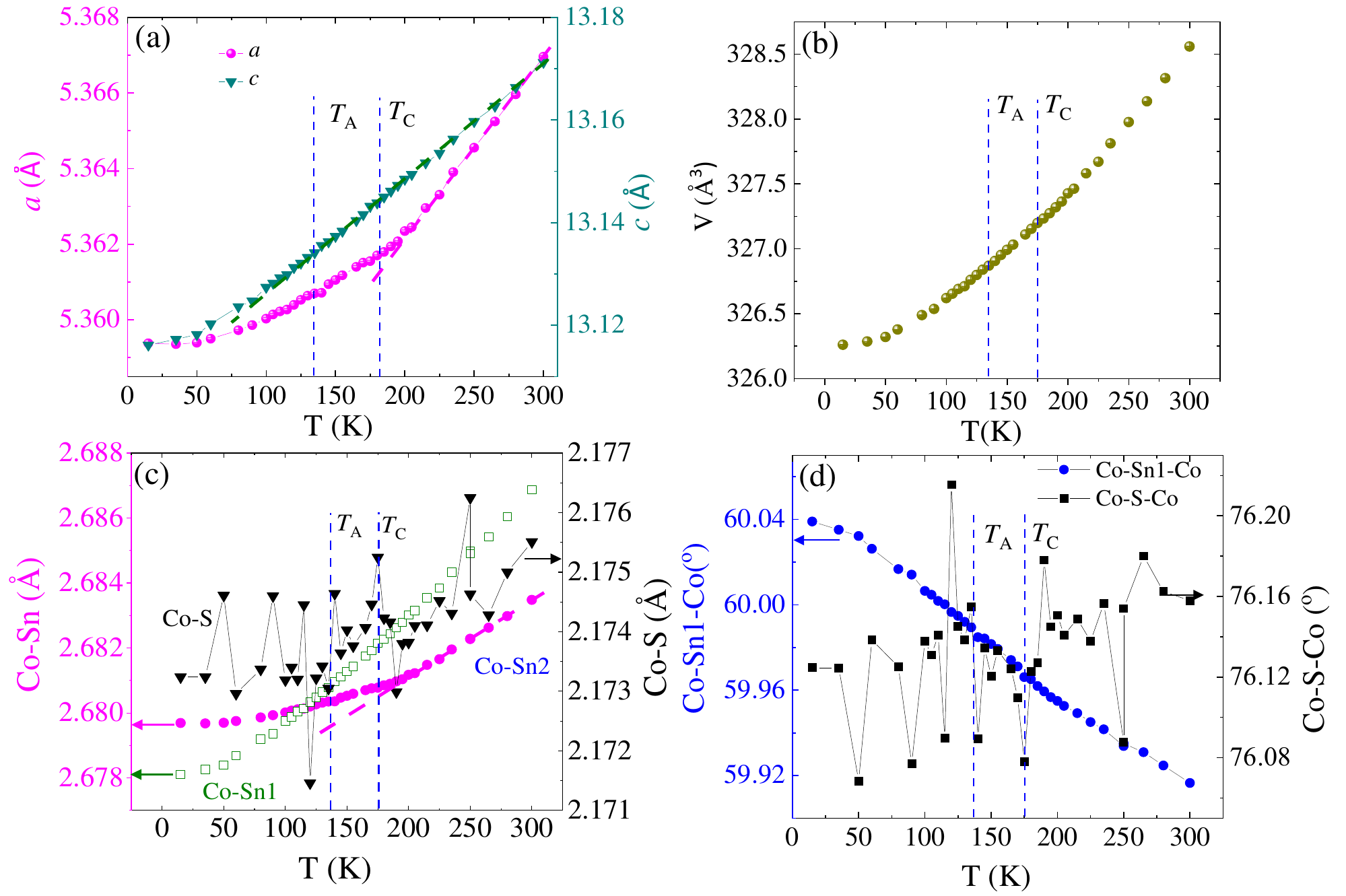}
\caption{(a) Temperature dependence of the lattice constants $a$ and $c$, (b) the volume of one unit cell, (c) the bond lengths Co-Sn(1), Co-Sn(2) and Co-S within one octahedra, and (d) the Co-Sn(1)-Co, Co-S-Co bond angles.   }
\label{fig:3}
\end{figure}

To explore the possible correlation between magnetic order and the crystal lattice, high resolution neutron powder diffraction experiments were employed to study the temperature dependence of the lattice constants, unit cell volume, bond distances and angles. The Rietveld analysis on the neutron diffraction patterns were shown in Fig. S1 (a-c) and the main results are displayed in Fig. 3(a-d). While the lattice constant $c$ exhibits a monotonous decrease without any anomaly at $T_{C}$ and $T_{A}$, there is a clear anomaly observed near $T_{C}$ in lattice constant $a$, unveiling a  spin-lattice coupling occurring in the $a_{H}b_{H}$ plane. The cobalt atoms are octahedrally coordinated by two S atoms and two in-equivalent Sn(1) and Sn(2) atoms. Within the octahedral environment of cobalt, the Co-S and Co-Sn(1) bond lengths, and the Co-S-Co and Co-Sn(1)-Co angles do not exhibit any anomaly at $T_{C}$ or $T_{A}$. The appearance of the ferromagnetic order drives a clear anomaly in the Co-Sn(2) bond lengths. As reported previously, the ferromagnetic order is dominated by the third-neighbor ``across-hexagon" $J_{d}$ model via Co-Sn(2)-Co exchange pathway\cite{Zhang2021}. Thus, this demonstrates an intimate connection between dominant $J_{d}$ and Co-Sn(2) bond lengths and the manifestation of the ferromagnetic order in  $T<T_{C}$.

\subsection{Local symmetry breaking from modeling neutron total scattering data} 

Motivated by the observed anomaly in lattice constant $a$ and Co-Sn(2)-Co length at $T_{C}$ from neutron diffraction, we switch to explore whether local distortion takes place by modeling high quality neutron total scattering data. Representative pair distribution functions (PDFs) going across the magnetic transition temperatures $T_{C}$ and $T_{A}$ are presented in Fig. 4(a). Interestingly, one can observe the emergence of new PDF peaks indicative of a local symmetry breaking as $T$ is lowered below $\approx$ $T_{C}$. Representative PDF patterns in a typical region $\approx$ 5 \AA~are depicted in the inset of Fig. 4(a). The detailed temperature dependence of the PDF patterns in the region of 4.2 \AA~$<$ r $<$5.55 \AA~shown in Fig. 2(f) indicates that strikingly, the new PDF peaks near 5.1 and 4.9 \AA~ co-emerge with the long-range FM order at $T_{C}$ and with the further decrease of $T$, they become stronger. The PDF peak intensity at 4.5 \AA~also shows a rapid increase below $T_{C}$. Note that the new PDF peak intensities near 5.1 and 4.9 \AA~ in Fig. 2(f) behaves like an order parameter of the long-range FM order shown in Fig. 2(e), which indicates that the local symmetry breaking emerges directly with the long-range FM order. 
 
\begin{figure*}
\centering
\includegraphics[width=1\linewidth]{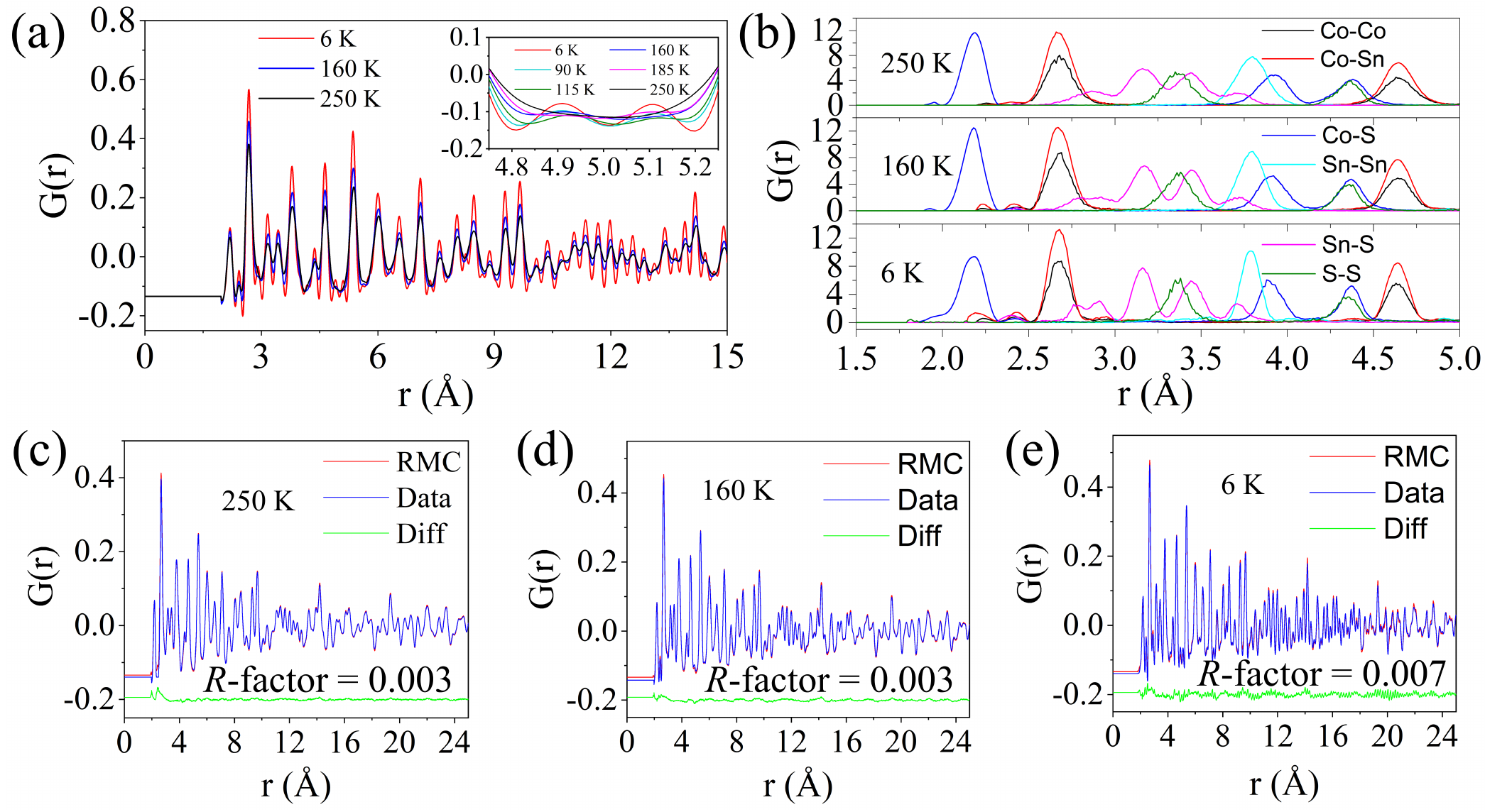}
\caption{ (a) Overview of PDFs at various temperature points and the insets are zoom-in over the regions around 5 \AA.  (b) Partial PDFs of all atomic pairs extracted from the fitted model corresponding to datasets at 250 K, 160 K and 6 K. (c), (d) and (e) shows the PDF fitting results for the datasets at 250 K, 160 K and 6 K, respectively - the $R$-factor of the fitting, as calculated by $\sum_i(y_i^o - y_i^c)^2/\sum_i(y_i^o)^2$ (where $y_i^o$ and $y_i^c$ represents the observed and calculated PDF intensity, respectively), is presented in each figure. 
   }
\label{fig:4}
\end{figure*}

Due to the very low moment, there is negligible magnetic contribution to either powder diffraction or PDF patterns here so that we can model the data without the need to consider the magnetic order. The RMC approach\cite{tucker:2007, yuanpeng:2020} we used here is different from the conventional crystallographic study in that it models the total scattering data (in both $Q$ and $r$-space) along with the Bragg data simultaneously and can unveil the rich information on the local structure within the average structural framework.  The fitting results on PDF patterns in real-space are presented in Fig. 4 (c-e), and the $Q$-space and Bragg results can be found in Fig. S2 in the supplementary material. The partial PDFs corresponding to each pair of atoms were calculated and put in Fig. 4 (b). Comparing among the three $T$ points, one can notice more complicated local bonding environments as the temperature is lowered to the magnetic ordered region. For example, the Sn-S partial exhibits a gradual splitting as $T$ decreases. Meanwhile, multiple partial PDFs show shoulder peaks around 2.5 \AA~in both 160 K and 6 K datasets, with a stronger signature at 6 K. Such features observed in multiple partial PDFs are consistent with the strengthening of the shoulder peak $\approx$2.5 \AA~between the first two main peaks in the experimental PDF patterns in Fig. 4 (a). Note that for the dataset above $T_{C}$, the shoulder peak $\approx$2.5 \AA~in PDF patterns can also be observed indicating that it comes from Fourier ripple, since no local distortion was found. Thus, the shoulder peak at 250 K in Fig. 4(a) can be treated as a baseline when considering the shoulder features $\approx$2.5 \AA~below $T_{C}$.

Although multiple partials (Co-Co, Co-Sn, etc.) exhibit shoulder features, it is still unclear in what form the local structure is distorted. Thus, the structural configurations obtained through RMC modeling were further analyzed. First, local bond angles were calculated and those Co and Sn(2) relevant angles in kagome lattice in the $a_{H}b_{H}$ plane are presented in Fig. 5 (a-b). For the 250 K dataset, both Co-Co-Co and Co-Sn(2)-Co angles show Gaussian-like distributions around the nominal position, i.e., 60$^\circ$ and 120$^\circ$. For the 160 K dataset, weak shoulder features emerge alongside both peaks for Co-Co-Co and Co-Sn(2)-Co and the signature becomes very clear for 6 K. Such features in both Co-Co-Co and Co-Sn(2)-Co angle distributions indicate that in $T<T_{C}$, the kagome lattice within the $a_{H}b_{H}$ plane is distorted locally breaking the local 3-fold point symmetry. Second, to further reveal the distortion of the kagome lattice and local atomic displacements, coordinates of all Co and Sn(2) residing in a single layer in the collapsed unit cell were projected onto the $a_{H}b_{H}$ plane, from which the atomic 
distribution density was estimated through kernel density estimation (KDE). Results at all the three temperatures are presented in Fig. 5 (c-e). The interesting finding here is an obvious skewed distribution of the cobalt kagome lattice that breaks three-fold symmetry as expected in rhombohedral structure. Similar signature of skewed distribution can also be observed in 160 K dataset but not in 250 K dataset. Such a skewed distribution for 6 K dataset is consistent with the emergence of shoulder peaks in the angle distributions. Third, to determine the local symmetry, we employed the program FINDSYM \cite{Stokes2005} to reconstruct the space group from the collapsed supercell configurations obtained via RMC modeling. Concerning the symmetry mining, a tolerance is set for the atomic displacement -- atoms displaced by less than the tolerance will be regarded as displacement-free. A higher tolerance tends to construct higher symmetry, and vice versa. A series of tolerance values were tried for all analysis from which a heatmap could be built as presented in Fig. 1(c). For 6 K dataset, the $Cm$ symmetry (No. 8) is determined for the tolerance as high as 0.055 \AA. The same low symmetry could also be observed at 160 K dataset with a much lower tolerance due to the weak distortion at 160 K.

Thus, upon cooling below $T_{C}$, although the global crystal symmetry is kept to be rhombohedral $R\mbox{-}3m$, there is a local symmetry breaking to monoclinic $Cm$. The relationship between these two structures is illustrated in Fig. 1(b). The obtained Wyckoff positions and average atomic positions over 12$\times$12$\times$6 supercell for the monoclinic structure at 6 K are summarized in Table 1. As compared to the average rhombohedral structure, one cobalt and sulfur site split into two due to the local atomic displacements in monoclinic $Cm$. As illustrated in Fig. 5 (e) at 6 K, the local atomic displacements of Co1 and Sn2 point to the +/- (100) direction in r.l.u of the rhombohedral structure, whereas the atomic displacements of Co2 atom exhibit distinct displacements and do not point to any high symmetric direction.  This can be understood well from the $Cm$ symmetry. As shown in the table I, the local displacements of Co1 and Sn2 occur in the $a_{m}c_{m}$ plane, and its projection in kagome lattice plane corresponds to the +/- (100) direction of the rhombohedral structure. As for Co2 atoms, the local displacements of Co2 occur in the 3-dimensional direction in the monoclinic cell and their projections in the kagome lattice plane do not point to any high symmetric direction. So there is a clear short-range (or local) lattice disorder with different atomic displacements and distributions for all the atoms within the monoclinic $Cm$ symmetry (the Fig. 5 (e) shows the atomic displacements and distributions for kagome-lattice Co atoms and central Sn2 atoms only). However, in the long-range scale, these local atomic displacements are canceled out, yielding rhombohedral structure in average below $T_{C}$. 

\begin{figure*}
\centering
\includegraphics[width=1\linewidth]{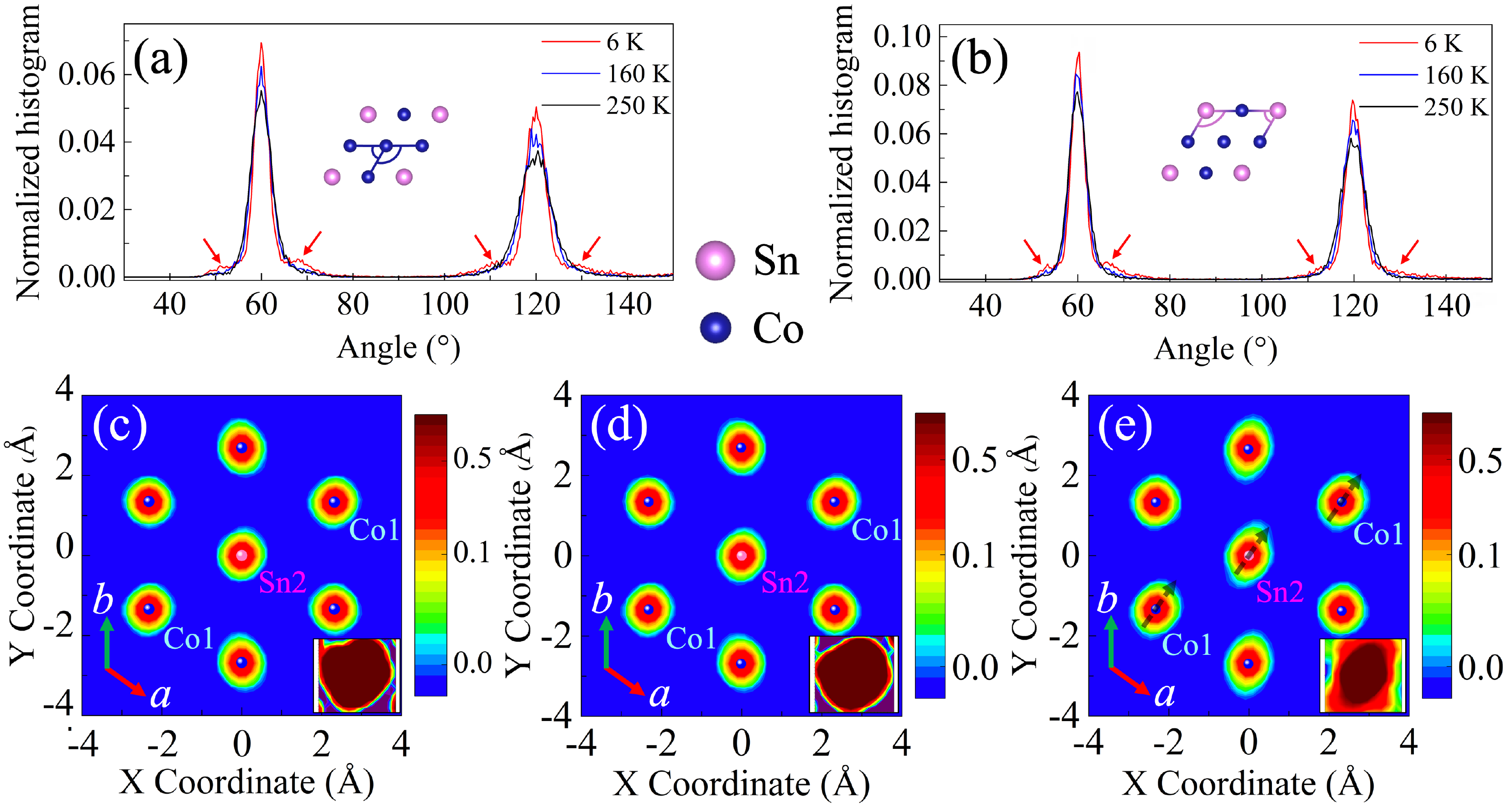}
\caption{ (a) \& (b) The Co-Co-Co and Co-Sn-Co bond angle extracted from the structural model obtained through RMC fitting to the datasets at 250, 160 and 6 K. The inset in each figure illustrates the corresponding bond angle, given the kagome lattice in the $a_{H}b_{H}$ plane (refer to the axis illustration on the left). (c), (d) and (e) is the KDE result for the atomic distribution for the cobalt kagome lattice and the central Sn2 atom in the collapsed unit cell at 250 K, 160 K and 6 K, respectively. Two Co1 and one Sn2 atoms are labeled and the four unlabeled atoms are Co2 atoms in monoclinic $Cm$ symmetry. The inset in each of (c-e) is a zoom-in of the KDE for the central Sn2 atom in the cobalt kagome lattice. The gray arrows indicate the the local atomic displacement direction of Co1 and Sn2 to the (100) direction in r.l.u. of the average rhombohedral structure.
   }
\label{fig:5}
\end{figure*}

\begin{table} 
\centering
\setlength{\abovecaptionskip}{10pt}%
\setlength{\belowcaptionskip}{10pt}%
\caption{Refined structural parameters for the local monoclinic symmetry with space group $Cm$ (No.8) at 6 K. The lattice constants are refined to be $a_{m}=$9.2752 \AA~, $b_{m}=$5.3594 \AA~ and $c_{m} =$5.3561 \AA~, $\beta=125.2^{o}$. }
\renewcommand{\arraystretch}{0.8}
\begin{tabular}{c|c|c|c|c}
\hline\hline
 atom & Wyckoff Site&  x & y &  z \\
 \hline
Co1 & 2a &   0.00162 &  0     &   0.50067    \\
Co2& 4b &   0.75136& 0.24971      &  0.00040        \\ 
Sn1 & 2a &    -0.00008 &  0.00000     &  -0.00243        \\
 Sn2 & 2a&    0.49945& 0.00000     &    0.50133       \\
 S1 & 2a &   0.71915  &  0.00000     &  0.28943        \\
 S2 & 2a&    0.28424& 0.00000     &   0.70960        \\
 \hline\hline  

\end{tabular}
\label{LowerSymmetry}
\end{table}

\subsection{DFT calculations on electronic band structures and Weyl points}

Next we investigate the effect of monoclinic lattice distortion on the magnetism and electronic band structure using density functional theory (DFT). Because of the difficulty including local distortion in the global undistorted structure, we consider the global structure in which the local distortion is repeated in the entire lattice. We first considered a FM state without spin-orbit coupling (SOC). Then, we turn on SOC and consider symmetry-allowed four FM states with Co ordered moments pointing along the $a_{m}$ direction (FM$_{a}$), the $b_{m}$ direction (FM$_{b}$), the $c_{m}$ direction (FM$_{c}$), and in-between $a_{m}$ and $c_{m}$ directions (FM$^{\ast}$) in the monoclinic notation. Note that the FM moment along the hexagonal c$_{H}$ axis is not allowed by the monoclinic symmetry $Cm$. Because the degree of lattice distortion is relatively small, the size of the ordered moment per Co is about 0.35 $\mu_{B}$ in all the cases: without SOC, FM$_{a}$, FM$_{b}$, and FM$^{\ast}$. Among all these four symmetry allowed FM states, FM$_{a}$ as shown in Fig. 6 (a) is found to be most stable, followed by FM$_{c}$ ($\approx$0.3 meV/f.u. higher than FM$_{a}$) and then FM$_{b}$ ($\approx$0.6 meV/f.u. higher than FM$_{a}$). The energy of FM$^{\ast}$ depends on the angle between the $a_{m}$ axis and the moment direction in the  $a_{m}c_{m}$ plane. It continuously increases from the energy of FM$_{a}$ at zero angle to the energy of FM$_{c}$ when the moment is along the $c_{m}$ axis. Given that there is only local monoclinic distortion in Co$_{3}$Sn$_{2}$S$_{2}$, this indicates that while the average long-range ferromagnetic moment points to the hexagonal $c_{H}$ axis, the local FM moment along the $c_{H}$ axis becomes unstable since it is forbidden by the local symmetry $Cm$. Instead, the local FM moment may deviate away from the $c_{H}$ axis and has the tendency to reorientate from the hexagonal $c_{H}$ axis to the monoclinic $a_{m}$ axis by $\approx$ 20$^{\rm o}$ toward the most stable FM$_{a}$ state, as illustrated in the right panel of Fig. 1(b).  Thus, the local symmetry breaking below $T_{C}$ is accompanied with the ferromagnetic instability, i.e., local changes in ferromagnetic moment direction.

\begin{figure*}
\centering
\includegraphics[width=0.95\linewidth]{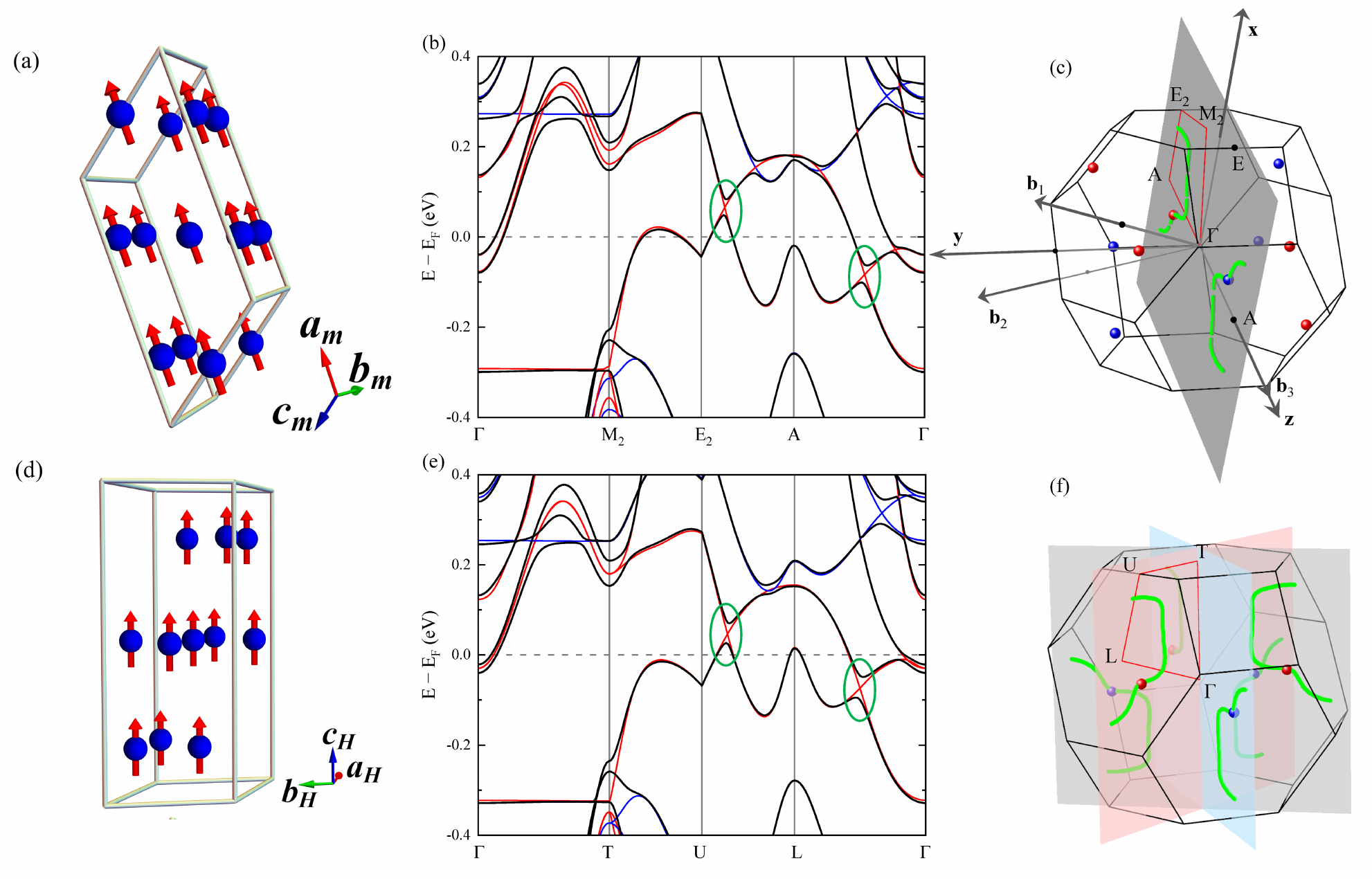}
\caption{(a) Magnetic structure and (b) electronic band structure of the most stable FM$_{a}$ phases in the distorted structure of Co$_{3}$Sn$_{2}$S$_{2}$. Red (blue) lines are majority (minority) spin bands without SOC, and black lines are results with SOC in the FM$_{a}$ phase, where Co moment is parallel to the $a_{m}$ axis. SOC lifts the degeneracy at the band crossing as shown as green circles. (c) First Brillouin zone of the distorted structure. Green lines indicate a nodal ring protected by the mirror symmetry without SOC. The mirror plane is indicated as a gray plane. Red (blue)
 spheres are Weyl points with positive (negative) chirality in the FM$_{a}$ phase. There are two Weyl points along the nodal ring. Reciprocal lattice vectors $\textbf{b}_{1,2,3}$ as well as Cartesian coordinates \textbf{x},\textbf{y},\textbf{z} are also indicated. Crystallographic axes $a$ and $b$ for the primitive cell are parallel to \textbf{x} and \textbf{y}, respectively. For comparison, panel (d-f) show the corresponding figures of undistorted structure. Three mirror planes are indicated by blue, pink and gray planes. }
\label{fig:6}
\end{figure*}

Figure 6(b) compares the electronic band structure of the most stable FM$_{a}$ order with moment along the monoclinic $a_{m}$ axis (see Fig. 6(a)) with and without SOC in the monoclinic distorted structure. Momentum is taken along high-symmetry lines as shown in Fig. 6(c). 
These band structures resemble those of the ground-state FM order along the $c_{H}$ axis (see Fig. 6 (d)) in the undistorted Co$_{3}$Sn$_{2}$S$_{2}$ as shown in Fig. 6(e), where the momentum is also taken along the corresponding high-symmetry lines in Fig. 6(f). Other FM states in the monoclinic distorted structure such as FM$_{b}$, FM$_{c}$, and FM$^{\ast}$ have similar band structures to that of FM$_{a}$ (shown in Fig S3 in the SM). To determine the existence and the location of Dirac points or nodal lines or Weyl points, we use the Wannier90 code \cite{Pizzi2020} and derive the maximally-localized Wannier functions. Detailed analysis was carried out using effective tight binding models with the WannierTools package \cite{WU2018405}. Similar to the band structure in the undistorted rhombohedral structure
\cite{Zhang2021}, there appear Dirac nodal lines without SOC. However, because of the monoclinic distortion, there is only one mirror plane (shown as a gray plane in Fig. 6(c)), resulting in one nodal ring shown by green lines in Fig.  6(c). SOC lifts the band degeneracy along this nodal line shown as green circles in Fig. 6(b) but keeps degeneracy at two points along the nodal ring. These two points form a pair of Weyl points. In the FM$_{a}$ phase, there appear additional 4 pairs of Weyl points that are not protected by the existing mirror symmetry. The existence and position of Weyl points are intimately related to the orientation of magnetic moment. Because the ordered moment is parallel to the mirror plane, the FM$_{a}$, FM$_{c}$ and FM$^{\ast}$ phases break the mirror symmetry but maintain mirror+time-reversal symmetry. Thus, these states have Weyl points along the nodal ring. On the other hand, since magnetic moment in the FM$_{b}$ state is perpendicular to the mirror plane, this state maintains the mirror symmetry. Therefore, the original nodal ring is protected in the FM$_{b}$ case (see Fig. S4 for details). In all cases, FM$_{a}$, FM$_{b}$, FM$_{c}$, and FM$^{\ast}$, there appear additional Weyl points away from the mirror plane. Results for  FM$_{b}$ and  FM$_{c}$ are presented in the Fig. S4 in the Supplemental Material.

\section{Discussion}
\subsection{Origin of local symmetry breaking}

The mismatch of local and average crystallographic structures establishes that Co$_{3}$Sn$_{2}$S$_{2}$ is an intrinsically lattice disordered system\cite{Allieta2012,Keen2015}. Interestingly, such mismatch in local and average symmetries only occurs below $T_{C}$ where the ferromagnetic order is formed, which is different from those in the nonmagnetic materials \cite{Mikkelse1982,EGAMI1991,Zhu2021} or  magnetic materials such as EuTiO$_{3}$\cite{Allieta2012} and Sr$_{1-x}$Na$_{x}$Fe$_{2}$As$_{2}$ (x=0.29 or 0.34)\cite{Frandsen2017}. In EuTiO$_{3}$, the local symmetry breaking occurs at 240 K, above the average lattice symmetry breaking temperature $\approx$ 235 K, and much higher than the AFM ordering temperature 5.3 K. In Sr$_{1-x}$Na$_{x}$Fe$_{2}$As$_{2}$ (x=0.29 or 0.34), the local symmetry breaking was found $\sim$ 100 K higher than the temperatures where both average lattice symmetry breaking and AFM transition occur simultaneously, i.e., 140 K for x=0.29 and 100 K for x=0.34. Furthermore, in both EuTiO$_{3}$ and Sr$_{1-x}$Na$_{x}$Fe$_{2}$As$_{2}$, the local symmetry breaking occurs at higher temperatures and at a lower temperature, the average lattice symmetry evolves to be the same to the local symmetry. Nevertheless, it does not happen to Co$_{3}$Sn$_{2}$S$_{2}$ and the competing local monoclinic and average rhombohedral symmetries persist from $T_{C}$ down to the lowest investigated temperature $\approx$ 6 K. In addition, as illustrated in Fig. 2(e) and (f), the local symmetry breaking in Co$_{3}$Sn$_{2}$S$_{2}$
exhibits a similar temperature dependence to the long-range FM signal. All of these results reveal a novel coupling between lattice and spin degrees of freedom in Co$_{3}$Sn$_{2}$S$_{2}$.

The geometric frustration\cite{Keen2015,Perversi2019} was proposed to be responsible for a local symmetry breaking in triangular and pyrochlore lattice compounds. Kagome lattice is a network of corner-sharing triangles, and there exists an intrinsic geometric frustration. Given that the local symmetry breaking emerges closely with the ferromagnetic order, we believe the intrinsic geometric frustration existing in the kagome lattice of cobalt and the ferromagnetic interactions in Co$_{3}$Sn$_{2}$S$_{2}$ are responsible for the appearance of local symmetry breaking below $T_{C}$.

\subsection{Relationship between local symmetry breaking and the puzzling magnetic properties}

The discovery of the mismatched local and average symmetries below $T_{C}$ provides new understandings on the puzzling magnetic properties in Co$_{3}$Sn$_{2}$S$_{2}$. The existence of the magnetic phase separation is incompatible with a simple rhombohedral structure with only one Co site. However, the local symmetry breaking from rhombohedral to monoclinic structures with local monoclinic distortion below $T_{C}$ allows the occurrence of separated FM and AFM phases in $T_{A}<T<T_{C}$. The mismatched local and average symmetries also provide an unambiguous experimental proof of the existence of local lattice/site disorder below $T_{C}$ in Co$_{3}$Sn$_{2}$S$_{2}$. Such local lattice/site disorder, unstable local ferromagnetic spins, and the geometric/magnetic frustration could well interpret the appearance of spin glass state below $T_{C}$. Upon cooling below $T_{A}$, the intrinsic lattice/site disorder becomes stronger, as revealed by stronger intensities of the new PDF peaks that have the tendency to saturate (see Fig. 2(f)). This could interpret why the spin glass state becomes stiffer below $T_{A}$ to pin the FM phase, leading to the exchange bias in both magnetic hysteresis loops and AHE. It is worthwhile mentioning that an interesting domain wall motion was observed at $T_{A}$ in Co$_{3}$Sn$_{2}$S$_{2}$ single crystal\cite{Lee2022}. Since the powder sample is not subject to the dynamics of domain walls, our total scattering measurements on powder sample allows to detect the intrinsic lattice/site disorder beyond the domain behavior.

The mismatched local and average symmetries suggest that the rhombohedral lattice becomes unstable, i.e., lattice instability/fluctuations below $T_{C}$, which is accompanied by the local ferromagnetic instability as discussed above. Such local ferromagnetic instability may reconcile the discrepancy between the saturation moment from the magnetization measurements\cite{Schnelle2013,Kassem2017,Zhang2021} and the average FM moment size derived from $\mu$ SR \cite{Guguchia2020} and our half polarized neutron diffraction measurements. The local FM moment tilting and instability tend to yield a lower average FM moment along $c_{H}$ than the saturation moment. 

\subsection{ Effect of the local monoclinic distortion on the Weyl property}
 
Our DFT calculations indicate that the consequences of local monoclinic distortion could be twofold to affect the Weyl properties: 1). Because two mirror planes out of three are lacking depending on the monoclinic axes, two pairs of Weyl points associated with those mirror symmetries are eliminated inside monoclinic regions. This could, in turn, induce finite broadening at bulk Weyl points. 2). At the boundary between a bulk hexagonal region and a local monoclinic region, or between monoclinic regions with different crystallographic axes, Fermi arcs could be formed because Weyl points in  monoclinic regions are gapped out. However, since the volume of each monoclinic region (or boundary area) is small, detecting such individual Fermi arc would be beyond the resolution of current measurements. Instead, broad topological surface band like feature such as broad Fermi arc may be formed in the bulk electronic structure. Such a feature could be detected by bulk sensitive probes, such as angle-resolved photoemission spectroscopy (ARPES). It comes to our attention that the experimental Fermi surface including Fermi arcs are broader than the $ab~initio$ calculation results using the average rhombohedral structure only as shown in Fig. 3 and 4(A) in Ref. \cite{Liu2019}. The local monoclinic distortion found here may contribute to understanding such difference. Advanced theoretical calculations on Fermi surface by considering both local symmetry $Cm$ and the global rhombohedral symmetry $R\mbox{-}3m$ are required to check such scenario below $T_{C}$ in Co$_{3}$Sn$_{2}$S$_{2}$. Further ARPES or scanning tunneling microscopy experiments are also needed to detect the detailed effect of the local monoclinic distortion on the Weyl properties in Co$_{3}$Sn$_{2}$S$_{2}$. 

A more accurate electronic structure by considering mismatched local and average symmetries are beyond the scope of our DFT capability since this requires a large supercell. Nevertheless, our DFT calculation demonstrates how the local monoclinic distortion could influence the electronic band structure and Weyl points by breaking mirror symmetries. In addition, due to the lattice instability between rhombohedral and monoclinic structures in Co$_{3}$Sn$_{2}$S$_{2}$, it would be of interest to explore a route such as chemical doping, high pressure, or thin film engineering to lower the average lattice symmetry to monoclinic $Cm$ for generating new magnetic and Weyl states as found by our DFT calculations. 

\subsection{Local symmetry breaking for engineering Co$_{3}$Sn$_{2}$S$_{2}$ and its derivatives}

The local symmetry deviated from the average symmetry has engineered attractive and unusual physical properties in a few functional materials, such as local structural dipoles in a cubic thermoelectric PbTe\cite{Bozin2010}, a tunable thermal expansion from negative to positive utilizing the local symmetry breaking of ScF$_{3}$\cite{Hu2018}, an enhanced $T_{C}$ associated with the lower local structure in FeSe$_{1-x}$Te$_{x}$ superconductors\cite{Louca2010}. In Co$_{3}$Sn$_{2}$S$_{2}$, the lattice instability manifested by mismatched local and average symmetries, and the magnetic instability indicate the existence of competing internal forces that are expected to be easily manipulated by a small external force such as composition, temperature(T), magnetic field $H$, pressure, thin film engineering to induce new physical properties. Indeed, many reports have shown that the magnetic, topological and thermoelectric properties in Co$_{3}$Sn$_{2}$S$_{2}$ are very sensitive to these external forces. The local structure should be considered in understanding these physical properties. For instance, Co$_{3}$Sn$_{2}$S$_{2}$ shows rich composition-$T$\cite{Guguchia2021,Shen2020} and $H-T$ phase diagrams\cite{Guguchia2020} involving distinct magnetic orders\cite{Guguchia2020,Guguchia2021,Shin2021}. The increase of the In-doping on Sn site reduces the FM fraction while promoting the in-plane AFM phase and destroys the topological state\cite{Guguchia2021}. This is probably associated with the enhancement of the monoclinic distortion by the indium doping in local scale and/or average scale. A magnetic field-induced skyrmion-like magnetic phase was also reported\cite{WU2020}. High pressure\cite{Liu2020} was found to suppress both ferromagnetic $T_{C}$ and AHE. As compared to single crystal form, Co$_{3}$Sn$_{2}$S$_{2}$ thin flakes\cite{Tanaka2020} exhibit enhanced electron mobility, large anomalous Hall conductivity and hall angle, whereas the large AHE disappeared below a critical thickness of $\sim$ 10 nm in Co$_{3}$Sn$_{2}$S$_{2}$ thin films\cite{Ikeda2021}. It deserves further investigation whether the local symmetry breaking still retains under such external forces and how the possible change in the local structure affects the physical properties. Given that both lattice and ferromagnetism are unstable below $T_{C}$ in Co$_{3}$Sn$_{2}$S$_{2}$, one may intently explore what and how external force could adjust the local symmetry or average symmetry for inducing novel magnetic, topological semimetallic and thermoelectric properties in Co$_{3}$Sn$_{2}$S$_{2}$ and its derivatives.

 \section{Conclusions and Outlook}

In summary, we discovered a hidden local symmetry breaking in the kagome-lattice Weyl semimetal Co$_{3}$Sn$_{2}$S$_{2}$ and unveiled the novel
correlation between structural complexity, magnetism and topological properties. While the global rhombohedral symmetry is unchanged, the geometric frustration in the kagome lattice of cobalt and the magnetic interactions drive a local monoclinic distortion involving a local kagome lattice distortion with anisotropic atomic displacements between two Co1 atoms and four Co2 atoms  below $T_{C}$. The competing local monoclinic and average rhombohedral structures retain to much lower temperatures without lowering average symmetry to be the same to the local symmetry. In addition, although the average ferromagnetic moment points to the hexagonal c$_{H}$ axis, the local monoclinic symmetry induces a local ferromagnetic moment canting away from this direction. These results indicate the existence of intrinsic lattice and magnetic instability below $T_{C}$. All of these features in Co$_{3}$Sn$_{2}$S$_{2}$ are unusual compared to other magnetic materials where local symmetry breaking was observed, showing a unique coupling between lattice and spin degree of freedom in Co$_{3}$Sn$_{2}$S$_{2}$. 

We have demonstrated that the mismatched local and average symmetries with the lattice disorder in Co$_{3}$Sn$_{2}$S$_{2}$ is a key in understanding previously puzzling appearance of magnetic phase separation and spin glass behavior. It also indicates that the previously puzzling magnetic properties were due to an incomplete understanding of the crystalline structure of Co$_{3}$Sn$_{2}$S$_{2}$. Moreover, the local monoclinic distortion plays a detrimental role in the formation of the Weyl points and may induce a broad topological surface band like feature from distorted regions by breaking mirror symmetries. Our study should motivate further theoretical and experimental studies to discover and control local structure for realizing new structure-property relationship and engineering desired physical properties more broadly in topological semimetals and kagome magnets. Total scattering is proved here to be one of the very powerful experimental tools. It also requires to pursue advanced theoretical calculations on the precise electronic structures and band topology on Co$_{3}$Sn$_{2}$S$_{2}$ and other topological materials, and to make a more careful interpretation on the experimental observation of their magnetic, electronic, and topological properties, by considering previously overlooked distinct local symmetries.

\section{Experimental}
 
\textbf{Sample preparation and half polarized neutron diffraction}

The preparation details of both polycrystalline and single crystalline samples were reported previously \cite{Zhang2021}. Two pieces of crystals were aligned in the ($HK0$) plane in the reciprocal lattice unit (r.l.u.) for the neutron experiments. The 1st half polarized neutron diffraction measurements were performed on the triple-axis neutron spectrometer HB-1 installed at the High Flux Isotope Reactor at Oak Ridge National Laboratory (ORNL), with a fixed incident neutron energy of 13.5 meV. For full polarization analysis, Heusler alloy (111) crystals were used as monochromator and analyzer. A flipping ratio of $\approx9$ was observed at nuclear reflections. For half polarization analysis, Heusler alloy (111) and Pyrolitic Graphite (002) crystals were used as monochromator and analyzer, respectively. A vertical magnetic field of 3 T cryomagnet was applied to point along the $c_{H}$ axis to align the ferromagnetic domains and saturate the magnetization of the sample.

To double check the ferromagnetic order and moment size, another half polarized neutron diffraction experiment on a different piece of crystal was conducted at time-of-flight instrument HYSPEC at the Spallation Neutron Source located in ORNL. The crystal was loaded inside a permanent-magnet yoke that provided a vertical magnetic field of $\sim$0.7 T, oriented parallel to the $c_{H}$ direction. An incident beam with the incident energy E$_{i}=$20meV was polarized using a Heusler monochromator, and a Mezei flipper was used to flip the spin direction of the incident neutron beam. The effective polarization factor of the incident neutron beam was estimated using the scattering from a Heusler single-crystal probe to be about 80 \%.

\textbf{Powder Neutron diffraction and pair distribution function (PDF) experiments}

High resolution neutron diffraction and pair distribution function (PDF) patterns were collected simultaneously using the neutron band with the center wavelength of 0.8 \AA~at powder diffractometer POWGEN, located at Spallation Neutron Source, Oak Ridge National Laboratory. Data collection was conducted at a series of temperatures from 6 K to 300 K. Rietveld analysis on the neutron diffraction data with a wide $Q$ coverage 1.3-16 \AA~was performed using GSAS-II package \cite{Toby:aj5212} to obtain the temperature dependence of the lattice constants, bond distances and angles. 

\textbf{Reverse Monte Carlo (RMC) analysis on neutron total scattering data}

 12$\times$12$\times$6 supercell configurations were built from the average structure corresponding to each temperature point and they were used as the starting configuration for RMC modeling. The RMC method relaxes structure configuration in a data-driven manner, following the Metropolis routine. For each temperature point, total scattering data in both $Q$ and $r$-space together with the Bragg scattering data were used for RMC modeling to provide constraints to the model. Here, the Bragg data for all temperature points were first refined with a Rietveld approach using Topas software, for which the RMCProfile package has an interface to read in peak profiles and background \cite{yuanpeng:2020}. Concerning the RMC modeling, the 12$\times$12$\times$6 supercell contains 18, 144 atoms and for 250 K and 160 K datasets, $\sim$300 moves per atom were tried among which $\sim$30 moves per atom were accepted on average. For the 6 K dataset, due to the complexity in local environment, the statistics are higher, given $\sim$900 moves per atom were tried among which $\sim$90 moves per atom were accepted. Given that there are multiple datasets involved in RMC modeling, an automatic weight assigning scheme was used\cite{tucker:2007}.

The structural distortion modes induced by irreducible representations of the parent  rhombohedral $R\mbox{-}3m$ space group and the symmetry-allowed magnetic structures for the monoclinic $Cm$ space group were analyzed using ISODISTORT\cite{Campbell2006}. The local symmetry $Cm$ in $T<T_{C}$ determined from modeling total scattering data is well consistent with the monoclinic distortion mode derived from the ISODISTORT.

\textbf{Density function theory (DFT) calculations}

Density function theory (DFT) calculations were carried out using the projector augmented wave method \cite{Kresse1999} with the generalized gradient approximation in the parametrization of Perdew, Burke, and Enzerhof \cite{Perdew1996} for exchange correlation as implemented in the Vienna ab initio simulation package (VASP) \cite{Kresse1996}. For Co and S standard potentials were used (Fe and S in the VASP distribution), and for Sn a potential, in which $d$ states were treated as valence states, is used (Sn$_{d}$). We use a $12\times12\times12$ $k$-point grid and an energy cutoff of 500 eV. The +U correction is not included because  Co$_{3}$Sn$_{2}$S$_{2}$ is an itinerant magnetic system. We use SeekK-path \cite{HINUMA2017140,Togo2018} to generate the primitive unit cell, that contained minimal number of atoms, i.e., three Co, two Sn and two S. 
 
\begin{acknowledgement}
The neutron research used resources at High Flux Isotope Reactor and Spallation Neutron Source, DOE Office of Science User Facilities operated by the Oak Ridge National Laboratory.  The research by SO, JY, MAM, DAT was sponsored by the Laboratory Directed Research and Development Program (LDRD) of Oak Ridge National Laboratory, managed by UT-Battelle, LLC, for the U.S. Department of Energy (Project ID 9533), with later stage supported by the US DOE, Office of Science, Basic Energy Sciences, Materials Sciences and Engineering Division (JY), and U.S. Department of Energy, Office of Science, National Quantum Information Science Research Centers, Quantum Science Center (SO, MAM, DAT).\\
\end{acknowledgement}

\begin{suppinfo}

Supporting Information includes the supplementary figures on the Rietveld analysis on the neutron diffraction and PDF patterns, as well as the supplementary DFT results.
The following files are available free of charge.
\begin{itemize}
  \item Filename:  Co$_{3}$Sn$_{2}$S$_{2}$-SM.pdf
 
\end{itemize}

\end{suppinfo}

\bibliography{References}

\providecommand{\latin}[1]{#1}
\makeatletter
\providecommand{\doi}
  {\begingroup\let\do\@makeother\dospecials
  \catcode`\{=1 \catcode`\}=2 \doi@aux}
\providecommand{\doi@aux}[1]{\endgroup\texttt{#1}}
\makeatother
\providecommand*\mcitethebibliography{\thebibliography}
\csname @ifundefined\endcsname{endmcitethebibliography}
  {\let\endmcitethebibliography\endthebibliography}{}
\begin{mcitethebibliography}{62}
\providecommand*\natexlab[1]{#1}
\providecommand*\mciteSetBstSublistMode[1]{}
\providecommand*\mciteSetBstMaxWidthForm[2]{}
\providecommand*\mciteBstWouldAddEndPuncttrue
  {\def\EndOfBibitem{\unskip.}}
\providecommand*\mciteBstWouldAddEndPunctfalse
  {\let\EndOfBibitem\relax}
\providecommand*\mciteSetBstMidEndSepPunct[3]{}
\providecommand*\mciteSetBstSublistLabelBeginEnd[3]{}
\providecommand*\EndOfBibitem{}
\mciteSetBstSublistMode{f}
\mciteSetBstMaxWidthForm{subitem}{(\alph{mcitesubitemcount})}
\mciteSetBstSublistLabelBeginEnd
  {\mcitemaxwidthsubitemform\space}
  {\relax}
  {\relax}

\bibitem[Egami and Billinge(2012)Egami, and Billinge]{Egami2012}
Egami,~T.; Billinge,~S.~J. Underneath the Bragg Peaks: Structural Analysis of
  Complex Materials. \emph{(Elsevier, Burlington, 2012).} \textbf{2012},
  \emph{2nd ed.}\relax
\mciteBstWouldAddEndPunctfalse
\mciteSetBstMidEndSepPunct{\mcitedefaultmidpunct}
{}{\mcitedefaultseppunct}\relax
\EndOfBibitem
\bibitem[Keen and Goodwin(2015)Keen, and Goodwin]{Keen2015}
Keen,~D.~A.; Goodwin,~A.~L. The crystallography of correlated disorder.
  \emph{Nature} \textbf{2015}, \emph{521}, 303--309\relax
\mciteBstWouldAddEndPuncttrue
\mciteSetBstMidEndSepPunct{\mcitedefaultmidpunct}
{\mcitedefaultendpunct}{\mcitedefaultseppunct}\relax
\EndOfBibitem
\bibitem[Zhu \latin{et~al.}(2021)Zhu, Huang, Ren, Zhang, Ke, Jen, Zhang, Wang,
  and Liu]{Zhu2021}
Zhu,~H.; Huang,~Y.; Ren,~J.; Zhang,~B.; Ke,~Y.; Jen,~A. K.-Y.; Zhang,~Q.;
  Wang,~X.-L.; Liu,~Q. Bridging Structural Inhomogeneity to Functionality: Pair
  Distribution Function Methods for Functional Materials Development.
  \emph{Advanced Science} \textbf{2021}, \emph{8}, 2003534\relax
\mciteBstWouldAddEndPuncttrue
\mciteSetBstMidEndSepPunct{\mcitedefaultmidpunct}
{\mcitedefaultendpunct}{\mcitedefaultseppunct}\relax
\EndOfBibitem
\bibitem[Wang \latin{et~al.}(2020)Wang, Zhao, Koch, Billinge, and
  Zunger]{Wang2020}
Wang,~Z.; Zhao,~X.-G.; Koch,~R.; Billinge,~S. J.~L.; Zunger,~A. {Understanding
  electronic peculiarities in tetragonal FeSe as local structural symmetry
  breaking}. \emph{Phys. Rev. B} \textbf{2020}, \emph{102}, 235121\relax
\mciteBstWouldAddEndPuncttrue
\mciteSetBstMidEndSepPunct{\mcitedefaultmidpunct}
{\mcitedefaultendpunct}{\mcitedefaultseppunct}\relax
\EndOfBibitem
\bibitem[Mikkelsen and Boyce(1982)Mikkelsen, and Boyce]{Mikkelse1982}
Mikkelsen,~J.~C.; Boyce,~J.~B. {Atomic-Scale Structure of Random Solid
  Solutions: Extended X-Ray-Absorption Fine-Structure Study of
  ${\mathrm{Ga}}_{1\ensuremath{-}x}{\mathrm{In}}_{x}\mathrm{As}$}. \emph{Phys.
  Rev. Lett.} \textbf{1982}, \emph{49}, 1412--1415\relax
\mciteBstWouldAddEndPuncttrue
\mciteSetBstMidEndSepPunct{\mcitedefaultmidpunct}
{\mcitedefaultendpunct}{\mcitedefaultseppunct}\relax
\EndOfBibitem
\bibitem[Egami \latin{et~al.}(1991)Egami, Rosenfeld, Toby, and A]{EGAMI1991}
Egami,~T.; Rosenfeld,~H.; Toby,~B.; A,~B. Diffraction studies of local
  atomic-structure in ferroelectric and superconducting oxides.
  \emph{Ferroelectrics} \textbf{1991}, \emph{120}, 11--21\relax
\mciteBstWouldAddEndPuncttrue
\mciteSetBstMidEndSepPunct{\mcitedefaultmidpunct}
{\mcitedefaultendpunct}{\mcitedefaultseppunct}\relax
\EndOfBibitem
\bibitem[Chong \latin{et~al.}(2012)Chong, Szczecinski, Bridges, Tucker,
  Claridge, and Rosseinsky]{Chong2012}
Chong,~S.~Y.; Szczecinski,~R.~J.; Bridges,~C.~A.; Tucker,~M.~G.;
  Claridge,~J.~B.; Rosseinsky,~M.~J. Local Structure of a Pure Bi A Site Polar
  Perovskite Revealed by Pair Distribution Function Analysis and Reverse Monte
  Carlo Modeling: Correlated Off-Axis Displacements in a Rhombohedral Material.
  \emph{J. Am. Chem. Soc.} \textbf{2012}, \emph{134}, 5836--5849\relax
\mciteBstWouldAddEndPuncttrue
\mciteSetBstMidEndSepPunct{\mcitedefaultmidpunct}
{\mcitedefaultendpunct}{\mcitedefaultseppunct}\relax
\EndOfBibitem
\bibitem[Bianconi \latin{et~al.}(1996)Bianconi, Saini, Lanzara, Missori,
  Rossetti, Oyanagi, Yamaguchi, Oka, and Ito]{Bianconi1996}
Bianconi,~A.; Saini,~N.~L.; Lanzara,~A.; Missori,~M.; Rossetti,~T.;
  Oyanagi,~H.; Yamaguchi,~H.; Oka,~K.; Ito,~T. {Determination of the Local
  Lattice Distortions in the Cu${\mathrm{O}}_{2}$ Plane of
  L${\mathrm{a}}_{1.85}$S${\mathrm{r}}_{0.15}$Cu${\mathrm{O}}_{4}$}.
  \emph{Phys. Rev. Lett.} \textbf{1996}, \emph{76}, 3412--3415\relax
\mciteBstWouldAddEndPuncttrue
\mciteSetBstMidEndSepPunct{\mcitedefaultmidpunct}
{\mcitedefaultendpunct}{\mcitedefaultseppunct}\relax
\EndOfBibitem
\bibitem[Louca \latin{et~al.}(1997)Louca, Egami, Brosha, R\"oder, and
  Bishop]{Despina1997}
Louca,~D.; Egami,~T.; Brosha,~E.~L.; R\"oder,~H.; Bishop,~A.~R. {Local
  Jahn-Teller distortion in
  ${\mathrm{La}}_{1\ensuremath{-}x}{\mathrm{Sr}}_{x}{\mathrm{MnO}}_{3}$
  observed by pulsed neutron diffraction}. \emph{Phys. Rev. B} \textbf{1997},
  \emph{56}, R8475--R8478\relax
\mciteBstWouldAddEndPuncttrue
\mciteSetBstMidEndSepPunct{\mcitedefaultmidpunct}
{\mcitedefaultendpunct}{\mcitedefaultseppunct}\relax
\EndOfBibitem
\bibitem[Perversi \latin{et~al.}(2019)Perversi, Pachoud, Cumby, Hudspeth,
  Wright, Kimber, and Attfield]{Perversi2019}
Perversi,~G.; Pachoud,~E.; Cumby,~J.; Hudspeth,~J.~M.; Wright,~J.~P.;
  Kimber,~S. A.~J.; Attfield,~J.~P. Co-emergence of magnetic order and
  structural fluctuations in magnetite. \emph{Nat. Commun.} \textbf{2019},
  \emph{10}, 2857\relax
\mciteBstWouldAddEndPuncttrue
\mciteSetBstMidEndSepPunct{\mcitedefaultmidpunct}
{\mcitedefaultendpunct}{\mcitedefaultseppunct}\relax
\EndOfBibitem
\bibitem[Frandsen \latin{et~al.}(2017)Frandsen, Taddei, Yi, Frano, Guguchia,
  Yu, Si, Bugaris, Stadel, Osborn, Rosenkranz, Chmaissem, and
  Birgeneau]{Frandsen2017}
Frandsen,~B.~A.; Taddei,~K.~M.; Yi,~M.; Frano,~A.; Guguchia,~Z.; Yu,~R.;
  Si,~Q.; Bugaris,~D.~E.; Stadel,~R.; Osborn,~R.; Rosenkranz,~S.;
  Chmaissem,~O.; Birgeneau,~R.~J. {Local Orthorhombicity in the Magnetic
  ${C}_{4}$ Phase of the Hole-Doped Iron-Arsenide Superconductor
  ${\mathrm{Sr}}_{1\ensuremath{-}x}{\mathrm{Na}}_{x}{\mathrm{Fe}}_{2}{\mathrm{As}}_{2}$}.
  \emph{Phys. Rev. Lett.} \textbf{2017}, \emph{119}, 187001\relax
\mciteBstWouldAddEndPuncttrue
\mciteSetBstMidEndSepPunct{\mcitedefaultmidpunct}
{\mcitedefaultendpunct}{\mcitedefaultseppunct}\relax
\EndOfBibitem
\bibitem[Jiang \latin{et~al.}(2021)Jiang, Bridges, Unocic, Pitike, Cooper,
  Zhang, Lin, and Page]{Jiang2021}
Jiang,~B.; Bridges,~C.~A.; Unocic,~R.~R.; Pitike,~K.~C.; Cooper,~V.~R.;
  Zhang,~Y.; Lin,~D.-Y.; Page,~K. Probing the Local Site Disorder and
  Distortion in Pyrochlore High-Entropy Oxides. \emph{J. Am. Chem. Soc.}
  \textbf{2021}, \emph{143}, 4193--4204\relax
\mciteBstWouldAddEndPuncttrue
\mciteSetBstMidEndSepPunct{\mcitedefaultmidpunct}
{\mcitedefaultendpunct}{\mcitedefaultseppunct}\relax
\EndOfBibitem
\bibitem[Goodwin \latin{et~al.}(2006)Goodwin, G., Dove, and Keen]{Goodwin2006}
Goodwin,~A.~L.; G.,~M.; Dove,~M.~T.; Keen,~D.~A. {Magnetic Structure of MnO at
  10 K from Total Neutron Scattering Data}. \emph{Phys. Rev. Lett.}
  \textbf{2006}, \emph{96}, 047209\relax
\mciteBstWouldAddEndPuncttrue
\mciteSetBstMidEndSepPunct{\mcitedefaultmidpunct}
{\mcitedefaultendpunct}{\mcitedefaultseppunct}\relax
\EndOfBibitem
\bibitem[Allieta \latin{et~al.}(2012)Allieta, Scavini, Spalek, Scagnoli,
  Walker, Panagopoulos, Saxena, Katsufuji, and Mazzoli]{Allieta2012}
Allieta,~M.; Scavini,~M.; Spalek,~L.~J.; Scagnoli,~V.; Walker,~H.~C.;
  Panagopoulos,~C.; Saxena,~S.~S.; Katsufuji,~T.; Mazzoli,~C. {Role of
  intrinsic disorder in the structural phase transition of magnetoelectric
  EuTiO${}_{3}$}. \emph{Phys. Rev. B} \textbf{2012}, \emph{85}, 184107\relax
\mciteBstWouldAddEndPuncttrue
\mciteSetBstMidEndSepPunct{\mcitedefaultmidpunct}
{\mcitedefaultendpunct}{\mcitedefaultseppunct}\relax
\EndOfBibitem
\bibitem[Lu \latin{et~al.}(2017)Lu, Song, Liu, Reyes, Kuhns, Lee, Fisher, and
  Mitrovic]{Lu2017}
Lu,~L.; Song,~M.; Liu,~W.; Reyes,~A.~P.; Kuhns,~P.; Lee,~H.~O.; Fisher,~I.~R.;
  Mitrovic,~V.~F. Magnetism and local symmetry breaking in a Mott insulator
  with strong spin orbit interactions. \emph{Nat. Commun.} \textbf{2017},
  \emph{8}, 14407\relax
\mciteBstWouldAddEndPuncttrue
\mciteSetBstMidEndSepPunct{\mcitedefaultmidpunct}
{\mcitedefaultendpunct}{\mcitedefaultseppunct}\relax
\EndOfBibitem
\bibitem[Trimarchi \latin{et~al.}(2018)Trimarchi, Wang, and
  Zunger]{Trimarchi2018}
Trimarchi,~G.; Wang,~Z.; Zunger,~A. {Polymorphous band structure model of
  gapping in the antiferromagnetic and paramagnetic phases of the Mott
  insulators MnO, FeO, CoO, and NiO}. \emph{Phys. Rev. B} \textbf{2018},
  \emph{97}, 035107\relax
\mciteBstWouldAddEndPuncttrue
\mciteSetBstMidEndSepPunct{\mcitedefaultmidpunct}
{\mcitedefaultendpunct}{\mcitedefaultseppunct}\relax
\EndOfBibitem
\bibitem[Park \latin{et~al.}(2011)Park, Lee, Wolff-Fabris, Koh, Eom, Kim,
  Farhan, Jo, Kim, Shim, and Kim]{Joonbum2011}
Park,~J.; Lee,~G.; Wolff-Fabris,~F.; Koh,~Y.~Y.; Eom,~M.~J.; Kim,~Y.~K.;
  Farhan,~M.~A.; Jo,~Y.~J.; Kim,~C.; Shim,~J.~H.; Kim,~J.~S. {Anisotropic Dirac
  Fermions in a Bi Square Net of ${\mathrm{SrMnBi}}_{2}$}. \emph{Phys. Rev.
  Lett.} \textbf{2011}, \emph{107}, 126402\relax
\mciteBstWouldAddEndPuncttrue
\mciteSetBstMidEndSepPunct{\mcitedefaultmidpunct}
{\mcitedefaultendpunct}{\mcitedefaultseppunct}\relax
\EndOfBibitem
\bibitem[Liu \latin{et~al.}({2017})Liu, Hu, Zhang, Graf, Cao, Radmanesh, Adams,
  Zhu, Cheng, Liu, Phelan, Wei, Jaime, Balakirev, Tennant, DiTusa, Chiorescu,
  Spinu, and Mao]{Liu2017}
Liu,~J.~Y. \latin{et~al.}  {A magnetic topological semimetal
  ${\mathrm{Sr}}_{1\ensuremath{-}y}{\mathrm{Mn}}_{1\ensuremath{-}z}{\mathrm{Sb}}_{2}$}.
  \emph{{Nat. Mater.}} \textbf{{2017}}, \emph{{16}}, {905}\relax
\mciteBstWouldAddEndPuncttrue
\mciteSetBstMidEndSepPunct{\mcitedefaultmidpunct}
{\mcitedefaultendpunct}{\mcitedefaultseppunct}\relax
\EndOfBibitem
\bibitem[Nakatsuji \latin{et~al.}({2015})Nakatsuji, Kiyohara, and
  Higo]{Nakatsuji2015}
Nakatsuji,~S.; Kiyohara,~N.; Higo,~T. {Large anomalous Hall effect in a
  non-collinear antiferromagnet at room temperature}. \emph{{Nature}}
  \textbf{{2015}}, \emph{{527}}, {212}\relax
\mciteBstWouldAddEndPuncttrue
\mciteSetBstMidEndSepPunct{\mcitedefaultmidpunct}
{\mcitedefaultendpunct}{\mcitedefaultseppunct}\relax
\EndOfBibitem
\bibitem[Nagaosa \latin{et~al.}(2020)Nagaosa, Morimoto, and
  Tokura]{Nagaosa2020}
Nagaosa,~N.; Morimoto,~T.; Tokura,~Y. Transport, magnetic and optical
  properties of Weyl Materials. \emph{Nat. Rev. Mater.} \textbf{2020},
  \emph{5}, 621--636\relax
\mciteBstWouldAddEndPuncttrue
\mciteSetBstMidEndSepPunct{\mcitedefaultmidpunct}
{\mcitedefaultendpunct}{\mcitedefaultseppunct}\relax
\EndOfBibitem
\bibitem[Savary and Balents(2016)Savary, and Balents]{Lucile2016}
Savary,~L.; Balents,~L. Quantum spin liquids: a review. \emph{Reports on
  Progress in Physics} \textbf{2016}, \emph{80}, 016502\relax
\mciteBstWouldAddEndPuncttrue
\mciteSetBstMidEndSepPunct{\mcitedefaultmidpunct}
{\mcitedefaultendpunct}{\mcitedefaultseppunct}\relax
\EndOfBibitem
\bibitem[Zhang \latin{et~al.}(2018)Zhang, Baker, Zhang, Wang, Wang, Su, Zhu,
  and Pratt]{Zhang2018}
Zhang,~B.; Baker,~P.~J.; Zhang,~Y.; Wang,~D.; Wang,~Z.; Su,~S.; Zhu,~D.;
  Pratt,~F.~L. Quantum Spin Liquid from a Three-Dimensional Copper-Oxalate
  Framework. \emph{J. Am. Chem. Soc.} \textbf{2018}, \emph{140}, 122--125\relax
\mciteBstWouldAddEndPuncttrue
\mciteSetBstMidEndSepPunct{\mcitedefaultmidpunct}
{\mcitedefaultendpunct}{\mcitedefaultseppunct}\relax
\EndOfBibitem
\bibitem[Yan and Felser(2017)Yan, and Felser]{Yan2017}
Yan,~B.; Felser,~C. Topological Materials: Weyl Semimetals. \emph{Annu. Rev.
  Condens. Matter Phys.} \textbf{2017}, \emph{8}, 337--354\relax
\mciteBstWouldAddEndPuncttrue
\mciteSetBstMidEndSepPunct{\mcitedefaultmidpunct}
{\mcitedefaultendpunct}{\mcitedefaultseppunct}\relax
\EndOfBibitem
\bibitem[Zou \latin{et~al.}({2019})Zou, He, and Xu]{Zou2019}
Zou,~J.; He,~Z.; Xu,~G. {The study of magnetic topological semimetals by first
  principles calculations}. \emph{{Npj Comput. Mater.}} \textbf{{2019}},
  \emph{{5}}, 96\relax
\mciteBstWouldAddEndPuncttrue
\mciteSetBstMidEndSepPunct{\mcitedefaultmidpunct}
{\mcitedefaultendpunct}{\mcitedefaultseppunct}\relax
\EndOfBibitem
\bibitem[Liu \latin{et~al.}({2018})Liu, Sun, Kumar, Muechler, Sun, Jiao, Yang,
  Liu, Liang, Xu, Kroder, Suess, Borrmann, Shekhar, Wang, Xi, Wang, Schnelle,
  Wirth, Chen, Goennenwein, and Felser]{Liu2018}
Liu,~E. \latin{et~al.}  {Giant anomalous Hall effect in a ferromagnetic
  kagome-lattice semimetal}. \emph{{ Nature Physics}} \textbf{{2018}},
  \emph{{14}}, {1125}\relax
\mciteBstWouldAddEndPuncttrue
\mciteSetBstMidEndSepPunct{\mcitedefaultmidpunct}
{\mcitedefaultendpunct}{\mcitedefaultseppunct}\relax
\EndOfBibitem
\bibitem[Liu \latin{et~al.}(2019)Liu, Liang, Liu, Xu, Li, Chen, Pei, Shi, Mo,
  Dudin, Kim, Cacho, Li, Sun, Yang, Liu, Parkin, Felser, and Chen]{Liu2019}
Liu,~D.~F. \latin{et~al.}  Magnetic Weyl semimetal phase in a Kagome crystal.
  \emph{Science} \textbf{2019}, \emph{365}, 1282--1285\relax
\mciteBstWouldAddEndPuncttrue
\mciteSetBstMidEndSepPunct{\mcitedefaultmidpunct}
{\mcitedefaultendpunct}{\mcitedefaultseppunct}\relax
\EndOfBibitem
\bibitem[Yin \latin{et~al.}({2019})Yin, Zhang, Chang, Wang, Tsirkin, Guguchia,
  Lian, Zhou, Jiang, Belopolski, Shumiya, Multer, Litskevich, Cochran, Lin,
  Wang, Neupert, Jia, Lei, and Hasan]{Yin2019}
Yin,~J.-X. \latin{et~al.}  {Negative flat band magnetism in a
  spin-orbit-coupled correlated kagome magnet}. \emph{{Nat. Phys.}}
  \textbf{{2019}}, \emph{{15}}, {443}\relax
\mciteBstWouldAddEndPuncttrue
\mciteSetBstMidEndSepPunct{\mcitedefaultmidpunct}
{\mcitedefaultendpunct}{\mcitedefaultseppunct}\relax
\EndOfBibitem
\bibitem[Morali \latin{et~al.}({2019})Morali, Batabyal, Nag, Liu, Xu, Sun, Yan,
  Felser, Avraham, and Beidenkopf]{Morali2019}
Morali,~N.; Batabyal,~R.; Nag,~P.~K.; Liu,~E.; Xu,~Q.; Sun,~Y.; Yan,~B.;
  Felser,~C.; Avraham,~N.; Beidenkopf,~H. {Fermi-arc diversity on surface
  terminations of the magnetic Weyl semimetal Co$_{3}$Sn$_{2}$S$_{2}$}.
  \emph{{Science}} \textbf{{2019}}, \emph{{365}}, {1286}\relax
\mciteBstWouldAddEndPuncttrue
\mciteSetBstMidEndSepPunct{\mcitedefaultmidpunct}
{\mcitedefaultendpunct}{\mcitedefaultseppunct}\relax
\EndOfBibitem
\bibitem[Yang \latin{et~al.}(2020)Yang, Zhang, Zhou, Dai, Liao, Weng, and
  Qiu]{Yang2020}
Yang,~R.; Zhang,~T.; Zhou,~L.; Dai,~Y.; Liao,~Z.; Weng,~H.; Qiu,~X.
  {Magnetization-Induced Band Shift in Ferromagnetic Weyl Semimetal
  ${\mathrm{Co}}_{3}{\mathrm{Sn}}_{2}{\mathrm{S}}_{2}$}. \emph{Phys. Rev.
  Lett.} \textbf{2020}, \emph{124}, 077403\relax
\mciteBstWouldAddEndPuncttrue
\mciteSetBstMidEndSepPunct{\mcitedefaultmidpunct}
{\mcitedefaultendpunct}{\mcitedefaultseppunct}\relax
\EndOfBibitem
\bibitem[Kassem \latin{et~al.}(2017)Kassem, Tabata, Waki, and
  Nakamura]{Kassem2017}
Kassem,~M.~A.; Tabata,~Y.; Waki,~T.; Nakamura,~H. {Low-field anomalous magnetic
  phase in The kagome-lattice shandite Sn$_{2}$Co$_{3}$S$_{2}$}. \emph{Phys.
  Rev. B} \textbf{2017}, \emph{96}, 014429\relax
\mciteBstWouldAddEndPuncttrue
\mciteSetBstMidEndSepPunct{\mcitedefaultmidpunct}
{\mcitedefaultendpunct}{\mcitedefaultseppunct}\relax
\EndOfBibitem
\bibitem[Schnelle \latin{et~al.}(2013)Schnelle, LeiThe-Jasper, Rosner,
  Schappacher, P\"ottgen, Pielnh~ofer, and Weihrich]{Schnelle2013}
Schnelle,~W.; LeiThe-Jasper,~A.; Rosner,~H.; Schappacher,~F.~M.; P\"ottgen,~R.;
  Pielnh~ofer,~F.; Weihrich,~R. {Ferromagnetic ordering and half-metallic state
  of Sn$_{2}$Co$_{3}$S$_{2}$ with The shandite-type structure}. \emph{Phys.
  Rev. B} \textbf{2013}, \emph{88}, 144404\relax
\mciteBstWouldAddEndPuncttrue
\mciteSetBstMidEndSepPunct{\mcitedefaultmidpunct}
{\mcitedefaultendpunct}{\mcitedefaultseppunct}\relax
\EndOfBibitem
\bibitem[Guguchia \latin{et~al.}({2020})Guguchia, Verezhak, Gawryluk, Tsirkin,
  Yin, Belopolski, Zhou, Simutis, Zhang, Cochran, Chang, Pomjakushina, Keller,
  Skrzeczkowska, Wang, Lei, Khasanov, Amato, Jia, Neupert, Luetkens, and
  Hasan]{Guguchia2020}
Guguchia,~Z. \latin{et~al.}  {Tunable anomalous Hall conductivity through
  volume-wise magnetic competition in a topological kagome magnet}. \emph{{Nat.
  Commun.}} \textbf{{2020}}, \emph{{11}}, 559\relax
\mciteBstWouldAddEndPuncttrue
\mciteSetBstMidEndSepPunct{\mcitedefaultmidpunct}
{\mcitedefaultendpunct}{\mcitedefaultseppunct}\relax
\EndOfBibitem
\bibitem[Lachman \latin{et~al.}({2020})Lachman, Murphy, Maksimovic, Kealh~ofer,
  Haley, McDonald, Long, and Analytis]{Lachman2020}
Lachman,~E.; Murphy,~R.~A.; Maksimovic,~N.; Kealh~ofer,~R.; Haley,~S.;
  McDonald,~R.~D.; Long,~J.~R.; Analytis,~J.~G. {Exchange biased anomalous Hall
  effect driven by frustration in a magnetic kagome lattice}. \emph{{ Nat.
  Commun.}} \textbf{{2020}}, \emph{{11}}, 560\relax
\mciteBstWouldAddEndPuncttrue
\mciteSetBstMidEndSepPunct{\mcitedefaultmidpunct}
{\mcitedefaultendpunct}{\mcitedefaultseppunct}\relax
\EndOfBibitem
\bibitem[Zhang \latin{et~al.}(2021)Zhang, Okamoto, Samolyuk, Stone, Kolesnikov,
  Xue, Yan, McGuire, Mandrus, and Tennant]{Zhang2021}
Zhang,~Q.; Okamoto,~S.; Samolyuk,~G.~D.; Stone,~M.~B.; Kolesnikov,~A.~I.;
  Xue,~R.; Yan,~J.; McGuire,~M.~A.; Mandrus,~D.; Tennant,~D.~A. {Unusual
  Exchange Couplings and Intermediate Temperature Weyl State in
  Co$_{3}$Sn$_{2}$S$_{2}$}. \emph{Phys. Rev. Lett.} \textbf{2021}, \emph{127},
  117201\relax
\mciteBstWouldAddEndPuncttrue
\mciteSetBstMidEndSepPunct{\mcitedefaultmidpunct}
{\mcitedefaultendpunct}{\mcitedefaultseppunct}\relax
\EndOfBibitem
\bibitem[Soh \latin{et~al.}(2022)Soh, Yi, Zivkovic, Qureshi, Stunault,
  Ouladdiaf, Rodr\'{\i}guez-Velamaz\'an, Shi, R\o{}nnow, and
  Boothroyd]{Soh2021}
Soh,~J.-R.; Yi,~C.; Zivkovic,~I.; Qureshi,~N.; Stunault,~A.; Ouladdiaf,~B.;
  Rodr\'{\i}guez-Velamaz\'an,~J.~A.; Shi,~Y.; R\o{}nnow,~H.~M.;
  Boothroyd,~A.~T. Magnetic structure of the topological semimetal
  ${\mathrm{Co}}_{3}{\mathrm{Sn}}_{2}{\mathrm{S}}_{2}$. \emph{Phys. Rev. B}
  \textbf{2022}, \emph{105}, 094435\relax
\mciteBstWouldAddEndPuncttrue
\mciteSetBstMidEndSepPunct{\mcitedefaultmidpunct}
{\mcitedefaultendpunct}{\mcitedefaultseppunct}\relax
\EndOfBibitem
\bibitem[Vaqueiro and Sobany(2009)Vaqueiro, and Sobany]{VAQUEIRO2009513}
Vaqueiro,~P.; Sobany,~G.~G. {A powder neutron diffraction study of The metallic
  ferromagnet Co$_{3}$Sn$_{2}$S$_{2}$}. \emph{Solid State Sciences}
  \textbf{2009}, \emph{11}, 513--518\relax
\mciteBstWouldAddEndPuncttrue
\mciteSetBstMidEndSepPunct{\mcitedefaultmidpunct}
{\mcitedefaultendpunct}{\mcitedefaultseppunct}\relax
\EndOfBibitem
\bibitem[Mydosh(1993)]{Mydosh1993}
Mydosh,~A.~J. \emph{{Spin Glasses: An Experimental Introduction}}; {Taylor and
  Francis}: London, 1993\relax
\mciteBstWouldAddEndPuncttrue
\mciteSetBstMidEndSepPunct{\mcitedefaultmidpunct}
{\mcitedefaultendpunct}{\mcitedefaultseppunct}\relax
\EndOfBibitem
\bibitem[Leli{\`{e}}vre-Berna \latin{et~al.}(2010)Leli{\`{e}}vre-Berna, Wills,
  Bourgeat-Lami, Dee, Hansen, Henry, Poole, Thomas, Tonon, Torregrossa,
  Andersen, Bordenave, Jullien, Mouveau, Gu{\'{e}}rard, and Manzin]{Leli2010}
Leli{\`{e}}vre-Berna,~E. \latin{et~al.}  Powder diffraction with spin polarized
  neutrons. \emph{Meas Sci Technol} \textbf{2010}, \emph{21}, 055106\relax
\mciteBstWouldAddEndPuncttrue
\mciteSetBstMidEndSepPunct{\mcitedefaultmidpunct}
{\mcitedefaultendpunct}{\mcitedefaultseppunct}\relax
\EndOfBibitem
\bibitem[Tucker \latin{et~al.}(2007)Tucker, Keen, Dove, Goodwin, and
  Hui]{tucker:2007}
Tucker,~M.~G.; Keen,~D.~A.; Dove,~M.~T.; Goodwin,~A.~L.; Hui,~Q. {RMCProfile:
  reverse Monte Carlo for polycrystalline Materials}. \emph{J. Condens. Matter
  Phys.} \textbf{2007}, \emph{19}, 335218\relax
\mciteBstWouldAddEndPuncttrue
\mciteSetBstMidEndSepPunct{\mcitedefaultmidpunct}
{\mcitedefaultendpunct}{\mcitedefaultseppunct}\relax
\EndOfBibitem
\bibitem[Zhang \latin{et~al.}(2020)Zhang, Eremenko, Krayzman, Tucker, and
  Levin]{yuanpeng:2020}
Zhang,~Y.; Eremenko,~M.; Krayzman,~V.; Tucker,~M.~G.; Levin,~I. New
  capabilities for enhancement of RMCPr ofile: instrumental pr ofiles with
  arbitrary peak shapes for structural refinements using The reverse Monte
  Carlo method. \emph{J. Appl. Cryst.} \textbf{2020}, \emph{53},
  1509--1518\relax
\mciteBstWouldAddEndPuncttrue
\mciteSetBstMidEndSepPunct{\mcitedefaultmidpunct}
{\mcitedefaultendpunct}{\mcitedefaultseppunct}\relax
\EndOfBibitem
\bibitem[Stokes and Hatch(2005)Stokes, and Hatch]{Stokes2005}
Stokes,~H.~T.; Hatch,~D.~M. {{\it FINDSYM}: program for identifying the
  space-group symmetry of a crystal}. \emph{J. Appl. Crystallogr.}
  \textbf{2005}, \emph{38}, 237--238\relax
\mciteBstWouldAddEndPuncttrue
\mciteSetBstMidEndSepPunct{\mcitedefaultmidpunct}
{\mcitedefaultendpunct}{\mcitedefaultseppunct}\relax
\EndOfBibitem
\bibitem[Pizzi \latin{et~al.}(2020)Pizzi, Vitale, Arita, Blügel, Freimuth,
  G{\'{e}}ranton, Gibertini, Gresch, Johnson, Koretsune, Iba{\~{n}}ez-Azpiroz,
  Lee, Lihm, Marchand, Marrazzo, Mokrousov, Mustafa, Nohara, Nomura, Paulatto,
  Ponc{\'{e}}, Ponweiser, Qiao, Thöle, Tsirkin, Wierzbowska, Marzari,
  Vanderbilt, Souza, ofi, and Yates]{Pizzi2020}
Pizzi,~G. \latin{et~al.}  Wannier90 as a community code: new features and
  applications. \emph{J. Condens. Matter Phys.} \textbf{2020}, \emph{32},
  165902\relax
\mciteBstWouldAddEndPuncttrue
\mciteSetBstMidEndSepPunct{\mcitedefaultmidpunct}
{\mcitedefaultendpunct}{\mcitedefaultseppunct}\relax
\EndOfBibitem
\bibitem[Wu \latin{et~al.}(2018)Wu, Zhang, Song, Troyer, and
  Soluyanov]{WU2018405}
Wu,~Q.; Zhang,~S.; Song,~H.-F.; Troyer,~M.; Soluyanov,~A.~A. WannierTools: An
  open-sources oftware package for novel topological Materials. \emph{Comput.
  Phys. Commun.} \textbf{2018}, \emph{224}, 405--416\relax
\mciteBstWouldAddEndPuncttrue
\mciteSetBstMidEndSepPunct{\mcitedefaultmidpunct}
{\mcitedefaultendpunct}{\mcitedefaultseppunct}\relax
\EndOfBibitem
\bibitem[Lee \latin{et~al.}(2022)Lee, Vir, Manna, Shekhar, Moore, Kastner,
  Felser, and Orenstein]{Lee2022}
Lee,~C.; Vir,~P.; Manna,~K.; Shekhar,~C.; Moore,~J.~E.; Kastner,~M.~A.;
  Felser,~C.; Orenstein,~J. {Observation of a phase transition within the
  domain walls of ferromagnetic Co$_{3}$Sn$_{2}$S$_{2}$}. \emph{Nat. Commun.}
  \textbf{2022}, \emph{13}, 3000\relax
\mciteBstWouldAddEndPuncttrue
\mciteSetBstMidEndSepPunct{\mcitedefaultmidpunct}
{\mcitedefaultendpunct}{\mcitedefaultseppunct}\relax
\EndOfBibitem
\bibitem[Bozin \latin{et~al.}(2010)Bozin, Malliakas, Souvatzis, Proffen,
  Spaldin, Kanatzidis, and Billinge]{Bozin2010}
Bozin,~E.~S.; Malliakas,~C.~D.; Souvatzis,~P.; Proffen,~T.; Spaldin,~N.~A.;
  Kanatzidis,~M.~G.; Billinge,~S. J.~L. Entropically Stabilized Local Dipole
  Formation in Lead Chalcogenides. \emph{Science} \textbf{2010}, \emph{330},
  1660--1663\relax
\mciteBstWouldAddEndPuncttrue
\mciteSetBstMidEndSepPunct{\mcitedefaultmidpunct}
{\mcitedefaultendpunct}{\mcitedefaultseppunct}\relax
\EndOfBibitem
\bibitem[Hu \latin{et~al.}(2018)Hu, Qin, Sanson, Huang, Pan, Li, Sun, Wang,
  Guo, Aydemir, Ren, Sun, Deng, Aquilanti, Rondinelli, Chen, and Xing]{Hu2018}
Hu,~L. \latin{et~al.}  {Localized Symmetry Breaking for Tuning Thermal
  Expansion in ScF$_{3}$ Nanoscale Frameworks}. \emph{J. Am. Chem. Soc.}
  \textbf{2018}, \emph{140}, 4477--4480\relax
\mciteBstWouldAddEndPuncttrue
\mciteSetBstMidEndSepPunct{\mcitedefaultmidpunct}
{\mcitedefaultendpunct}{\mcitedefaultseppunct}\relax
\EndOfBibitem
\bibitem[Louca \latin{et~al.}(2010)Louca, Horigane, Llobet, Arita, Ji,
  Katayama, Konbu, Nakamura, Koo, Tong, and Yamada]{Louca2010}
Louca,~D.; Horigane,~K.; Llobet,~A.; Arita,~R.; Ji,~S.; Katayama,~N.;
  Konbu,~S.; Nakamura,~K.; Koo,~T.-Y.; Tong,~P.; Yamada,~K. {Local atomic
  structure of superconducting
  ${\text{FeSe}}_{1\ensuremath{-}x}{\text{Te}}_{x}$}. \emph{Phys. Rev. B}
  \textbf{2010}, \emph{81}, 134524\relax
\mciteBstWouldAddEndPuncttrue
\mciteSetBstMidEndSepPunct{\mcitedefaultmidpunct}
{\mcitedefaultendpunct}{\mcitedefaultseppunct}\relax
\EndOfBibitem
\bibitem[Guguchia \latin{et~al.}(2021)Guguchia, Zhou, Wang, Yin, Mielke,
  Tsirkin, Belopolski, Zhang, Cochran, Neupert, Khasanov, Amato, Jia, Hasan,
  and Luetkens]{Guguchia2021}
Guguchia,~Z.; Zhou,~H.; Wang,~C.~N.; Yin,~J.-X.; Mielke,~C.; Tsirkin,~S.~S.;
  Belopolski,~I.; Zhang,~S.-S.; Cochran,~T.~A.; Neupert,~T.; Khasanov,~R.;
  Amato,~A.; Jia,~S.; Hasan,~M.~Z.; Luetkens,~H. {Multiple quantum phase
  transitions of different nature in the topological kagome magnet
  Co$_{3}$Sn$_{2-x}$$In$$_{x}$S$_{2}$ }. \emph{Npj quantum mater}
  \textbf{2021}, \emph{6}, 50\relax
\mciteBstWouldAddEndPuncttrue
\mciteSetBstMidEndSepPunct{\mcitedefaultmidpunct}
{\mcitedefaultendpunct}{\mcitedefaultseppunct}\relax
\EndOfBibitem
\bibitem[Shen \latin{et~al.}(2020)Shen, Zeng, Zhang, Sun, Yao, Xi, Wang, Wu,
  Shen, Liu, and Liu]{Shen2020}
Shen,~J.; Zeng,~Q.; Zhang,~S.; Sun,~H.; Yao,~Q.; Xi,~X.; Wang,~W.; Wu,~G.;
  Shen,~B.; Liu,~Q.; Liu,~E. {33\% Giant Anomalous Hall Current Driven by Both
  Intrinsic and Extrinsic Contributions in Magnetic Weyl Semimetal
  Co$_{3}$Sn$_{2}$S$_{2}$}. \emph{Adv. Funct. Mater.} \textbf{2020}, \emph{30},
  2000830\relax
\mciteBstWouldAddEndPuncttrue
\mciteSetBstMidEndSepPunct{\mcitedefaultmidpunct}
{\mcitedefaultendpunct}{\mcitedefaultseppunct}\relax
\EndOfBibitem
\bibitem[Shin \latin{et~al.}(accessed 2021-05-09)Shin, Jun, Lee, and
  Jung]{Shin2021}
Shin,~D.-H.; Jun,~J.-H.; Lee,~S.-E.; Jung,~M.-H. Degenerate magnetic ground
  state and metastable state on trihexagonal Co-sublattices in
  Co$_{3}$Sn$_{2}$(S,Se)$_{2}$ single crystals. accessed 2021-05-09;
  \url{https://arxiv.org/abs/2105.03892}\relax
\mciteBstWouldAddEndPuncttrue
\mciteSetBstMidEndSepPunct{\mcitedefaultmidpunct}
{\mcitedefaultendpunct}{\mcitedefaultseppunct}\relax
\EndOfBibitem
\bibitem[Wu \latin{et~al.}(2020)Wu, Sun, Hsieh, Chen, Kakarla, Deng, Chu, and
  Yang]{WU2020}
Wu,~H.; Sun,~P.; Hsieh,~D.; Chen,~H.; Kakarla,~D.~C.; Deng,~L.; Chu,~C.;
  Yang,~H. {Observation of skyrmion-like magnetism in magnetic Weyl semimetal
  Co$_{3}$Sn$_{2}$S$_{2}$}. \emph{Mater. Today Phys.} \textbf{2020}, \emph{12},
  100189\relax
\mciteBstWouldAddEndPuncttrue
\mciteSetBstMidEndSepPunct{\mcitedefaultmidpunct}
{\mcitedefaultendpunct}{\mcitedefaultseppunct}\relax
\EndOfBibitem
\bibitem[Liu \latin{et~al.}(2020)Liu, Zhang, Xu, Yang, Wang, Lei, Sui, Uwatoko,
  Wang, Weng, Sun, and Cheng]{Liu2020}
Liu,~Z.~Y.; Zhang,~T.; Xu,~S.~X.; Yang,~P.~T.; Wang,~Q.; Lei,~H.~C.; Sui,~Y.;
  Uwatoko,~Y.; Wang,~B.~S.; Weng,~H.~M.; Sun,~J.~P.; Cheng,~J.-G. {Pressure
  effect on the anomalous Hall effect of ferromagnetic Weyl semimetal
  Co$_{3}$Sn$_{2}$S$_{2}$}. \emph{Phys. Rev. Mater.} \textbf{2020}, \emph{4},
  044203\relax
\mciteBstWouldAddEndPuncttrue
\mciteSetBstMidEndSepPunct{\mcitedefaultmidpunct}
{\mcitedefaultendpunct}{\mcitedefaultseppunct}\relax
\EndOfBibitem
\bibitem[Tanaka \latin{et~al.}(2020)Tanaka, Fujishiro, Mogi, Kaneko, Yokosawa,
  Kanazawa, Minami, Koretsune, Arita, Tarucha, Yamamoto, and
  Tokura]{Tanaka2020}
Tanaka,~M.; Fujishiro,~Y.; Mogi,~M.; Kaneko,~Y.; Yokosawa,~T.; Kanazawa,~N.;
  Minami,~S.; Koretsune,~T.; Arita,~R.; Tarucha,~S.; Yamamoto,~M.; Tokura,~Y.
  {Topological Kagome Magnet Co$_{3}$Sn$_{2}$S$_{2}$ Thin Flakes with High
  Electron Mobility and Large Anomalous Hall Effect}. \emph{Nano Lett.}
  \textbf{2020}, \emph{20}, 7476--7481\relax
\mciteBstWouldAddEndPuncttrue
\mciteSetBstMidEndSepPunct{\mcitedefaultmidpunct}
{\mcitedefaultendpunct}{\mcitedefaultseppunct}\relax
\EndOfBibitem
\bibitem[Ikeda \latin{et~al.}(2021)Ikeda, Fujiwara, Shiogai, Seki, Nomura,
  Takanashi, and Tsukazaki]{Ikeda2021}
Ikeda,~J.; Fujiwara,~K.; Shiogai,~J.; Seki,~T.; Nomura,~K.; Takanashi,~K.;
  Tsukazaki,~A. {Critical thickness for the emergence of Weyl features in
  Co$_{3}$Sn$_{2}$S$_{2}$ thin films}. \emph{Commun. Mater} \textbf{2021},
  \emph{2}, 18\relax
\mciteBstWouldAddEndPuncttrue
\mciteSetBstMidEndSepPunct{\mcitedefaultmidpunct}
{\mcitedefaultendpunct}{\mcitedefaultseppunct}\relax
\EndOfBibitem
\bibitem[Toby and Von~Dreele(2013)Toby, and Von~Dreele]{Toby:aj5212}
Toby,~B.~H.; Von~Dreele,~R.~B. {{\it GSAS-II}: The genesis of a modern
  open-source all purpose crystallography s oftware package}. \emph{J. Appl.
  Crystallogr.} \textbf{2013}, \emph{46}, 544--549\relax
\mciteBstWouldAddEndPuncttrue
\mciteSetBstMidEndSepPunct{\mcitedefaultmidpunct}
{\mcitedefaultendpunct}{\mcitedefaultseppunct}\relax
\EndOfBibitem
\bibitem[B.~J.~Campbell and Hatchl(2006)B.~J.~Campbell, and
  Hatchl]{Campbell2006}
B.~J.~Campbell,~D. E.~T.,~H. T.~Stokes; Hatchl,~D.~M. ISODISPLACE: An Internet
  Tool for Exploring Structural Distortions. \emph{J. Appl. Cryst.}
  \textbf{2006}, \emph{39}, 607--614\relax
\mciteBstWouldAddEndPuncttrue
\mciteSetBstMidEndSepPunct{\mcitedefaultmidpunct}
{\mcitedefaultendpunct}{\mcitedefaultseppunct}\relax
\EndOfBibitem
\bibitem[Kresse and Joubert(1999)Kresse, and Joubert]{Kresse1999}
Kresse,~G.; Joubert,~D. From ultras oft pseudopotentials to The projector
  augmented-wave method. \emph{Phys. Rev. B} \textbf{1999}, \emph{59},
  1758--1775\relax
\mciteBstWouldAddEndPuncttrue
\mciteSetBstMidEndSepPunct{\mcitedefaultmidpunct}
{\mcitedefaultendpunct}{\mcitedefaultseppunct}\relax
\EndOfBibitem
\bibitem[Perdew \latin{et~al.}(1996)Perdew, Burke, and Ernzerh~of]{Perdew1996}
Perdew,~J.~P.; Burke,~K.; Ernzerh~of,~M. Generalized Gradient Approximation
  Made Simple. \emph{Phys. Rev. Lett.} \textbf{1996}, \emph{77},
  3865--3868\relax
\mciteBstWouldAddEndPuncttrue
\mciteSetBstMidEndSepPunct{\mcitedefaultmidpunct}
{\mcitedefaultendpunct}{\mcitedefaultseppunct}\relax
\EndOfBibitem
\bibitem[Kresse and Furthm\"uller(1996)Kresse, and Furthm\"uller]{Kresse1996}
Kresse,~G.; Furthm\"uller,~J. Efficient iterative schemes for ab initio
  total-energy calculations using a plane-wave basis set. \emph{Phys. Rev. B}
  \textbf{1996}, \emph{54}, 11169--11186\relax
\mciteBstWouldAddEndPuncttrue
\mciteSetBstMidEndSepPunct{\mcitedefaultmidpunct}
{\mcitedefaultendpunct}{\mcitedefaultseppunct}\relax
\EndOfBibitem
\bibitem[Hinuma \latin{et~al.}(2017)Hinuma, Pizzi, Kumagai, Oba, and
  Tanaka]{HINUMA2017140}
Hinuma,~Y.; Pizzi,~G.; Kumagai,~Y.; Oba,~F.; Tanaka,~I. Band structure diagram
  paths based on crystallography. \emph{Comput. Mater. Sci.} \textbf{2017},
  \emph{128}, 140--184\relax
\mciteBstWouldAddEndPuncttrue
\mciteSetBstMidEndSepPunct{\mcitedefaultmidpunct}
{\mcitedefaultendpunct}{\mcitedefaultseppunct}\relax
\EndOfBibitem
\bibitem[Togo and Tanaka(accessed 2018-08-05)Togo, and Tanaka]{Togo2018}
Togo,~A.; Tanaka,~I. $\texttt{Spglib}$: a software library for crystal symmetry
  search. accessed 2018-08-05; \url{https://arxiv.org/abs/1808.01590}\relax
\mciteBstWouldAddEndPuncttrue
\mciteSetBstMidEndSepPunct{\mcitedefaultmidpunct}
{\mcitedefaultendpunct}{\mcitedefaultseppunct}\relax
\EndOfBibitem
\end{mcitethebibliography}

%
%
%
%
%
\end{document}


%

\begin{figure*}
\centering
\includegraphics[width=0.6\linewidth]{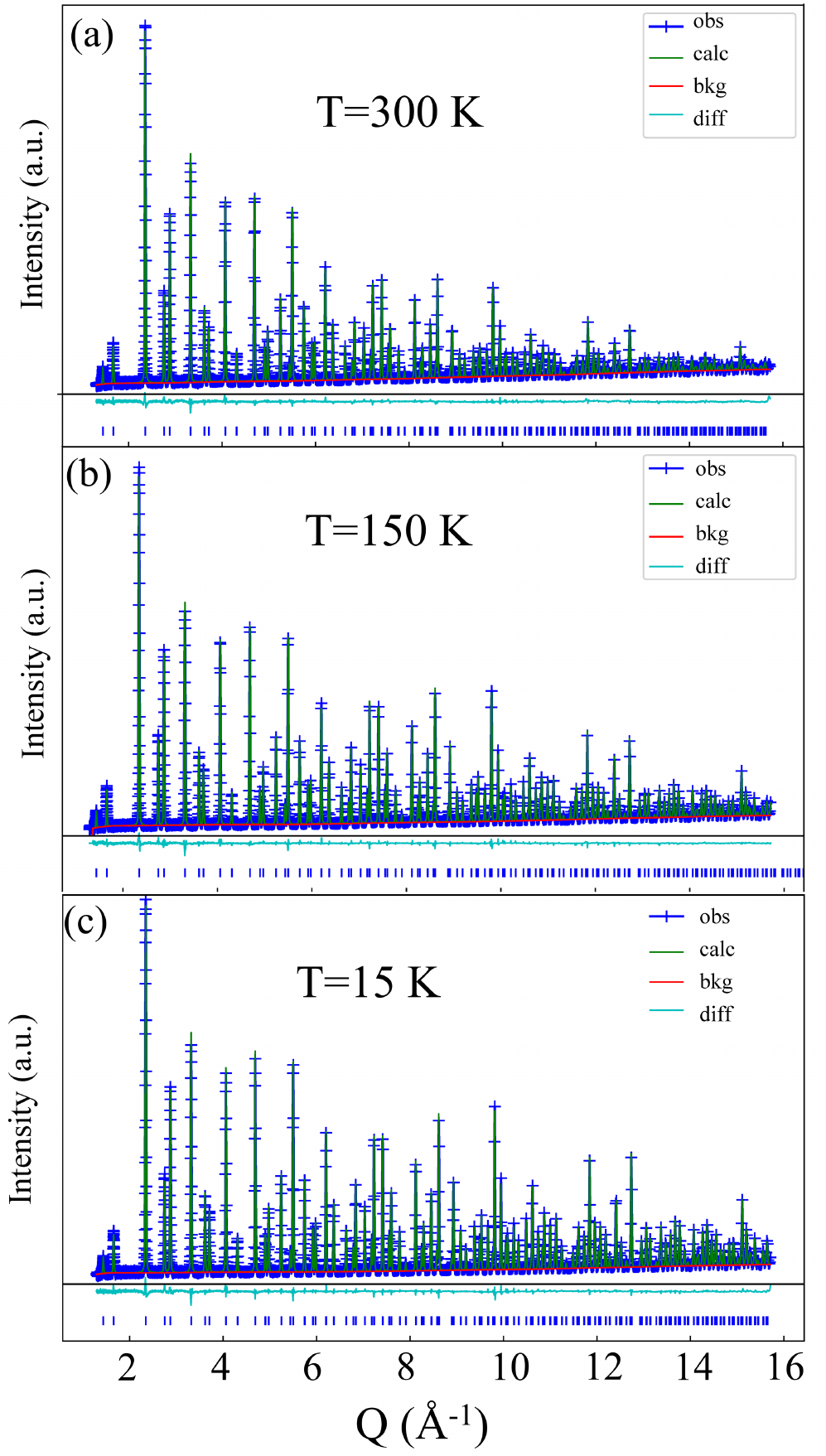}
\caption{Rietveld analysis on the high resolution neutron diffraction patterns in a wide Q region at (a) 300 K, (b) 150 K and (c) 15 K}
\label{fig:S1}
\end{figure*}

\begin{figure*}
\centering
\includegraphics[width=1\linewidth]{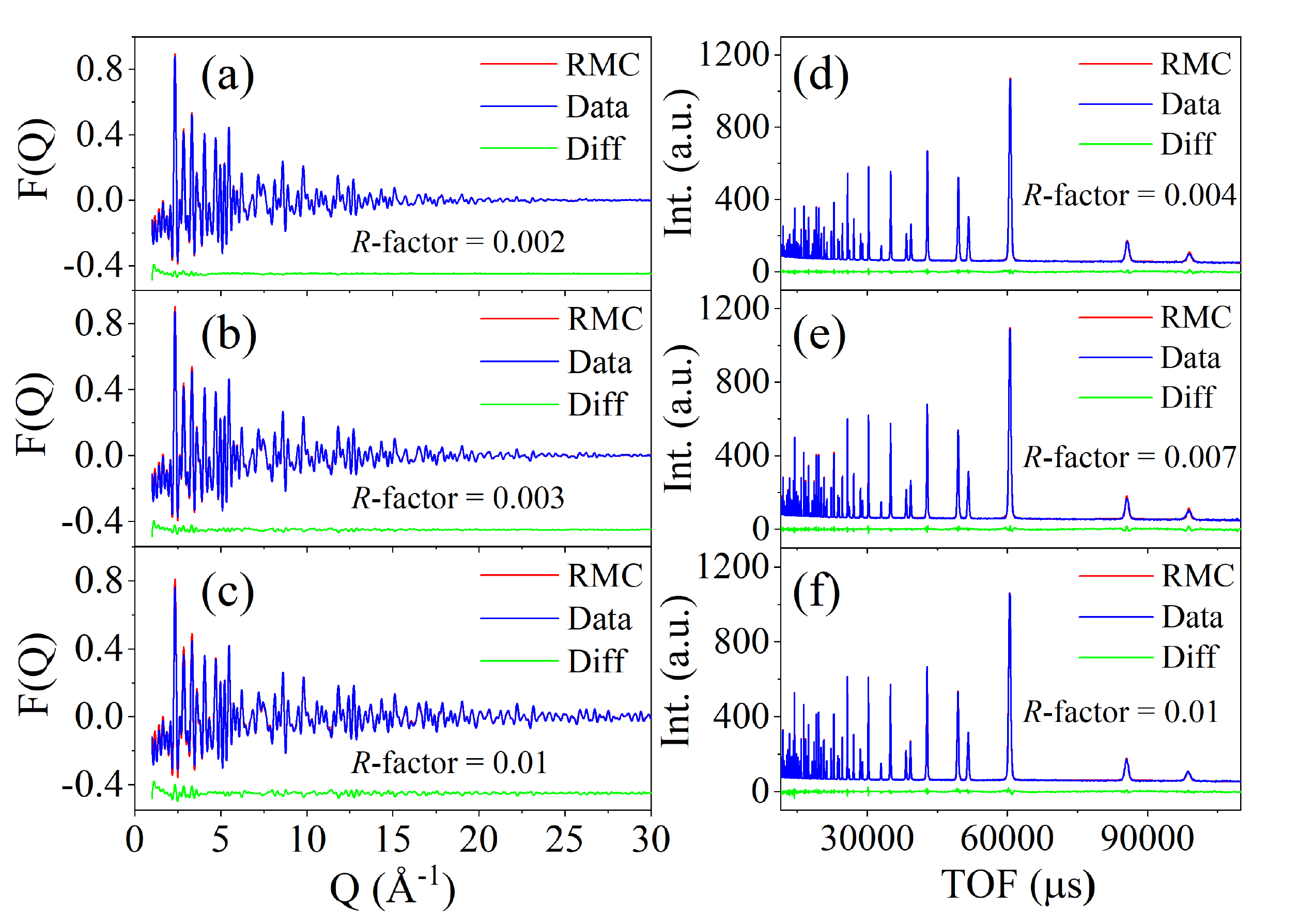}
\caption{(a-c) The F(Q) fitting results for datasets corresponding to 250 K, 160 K and 6 K, respectively. (d-f) The Bragg fitting results for datasets corresponding to 250 K, 160 K and 6 K, respectively. The $R$-factor of the fitting, as calculated by $\sum_i(y_i^o - y_i^c)^2/\sum_i(y_i^o)^2$ (where $y_i^o$ and $y_i^c$ represents the observed and calculated PDF intensity, respectively), is presented in each figure.}
\label{fig:SX1}
\end{figure*}

\begin{figure*}
\centering
\includegraphics[width=1\linewidth]{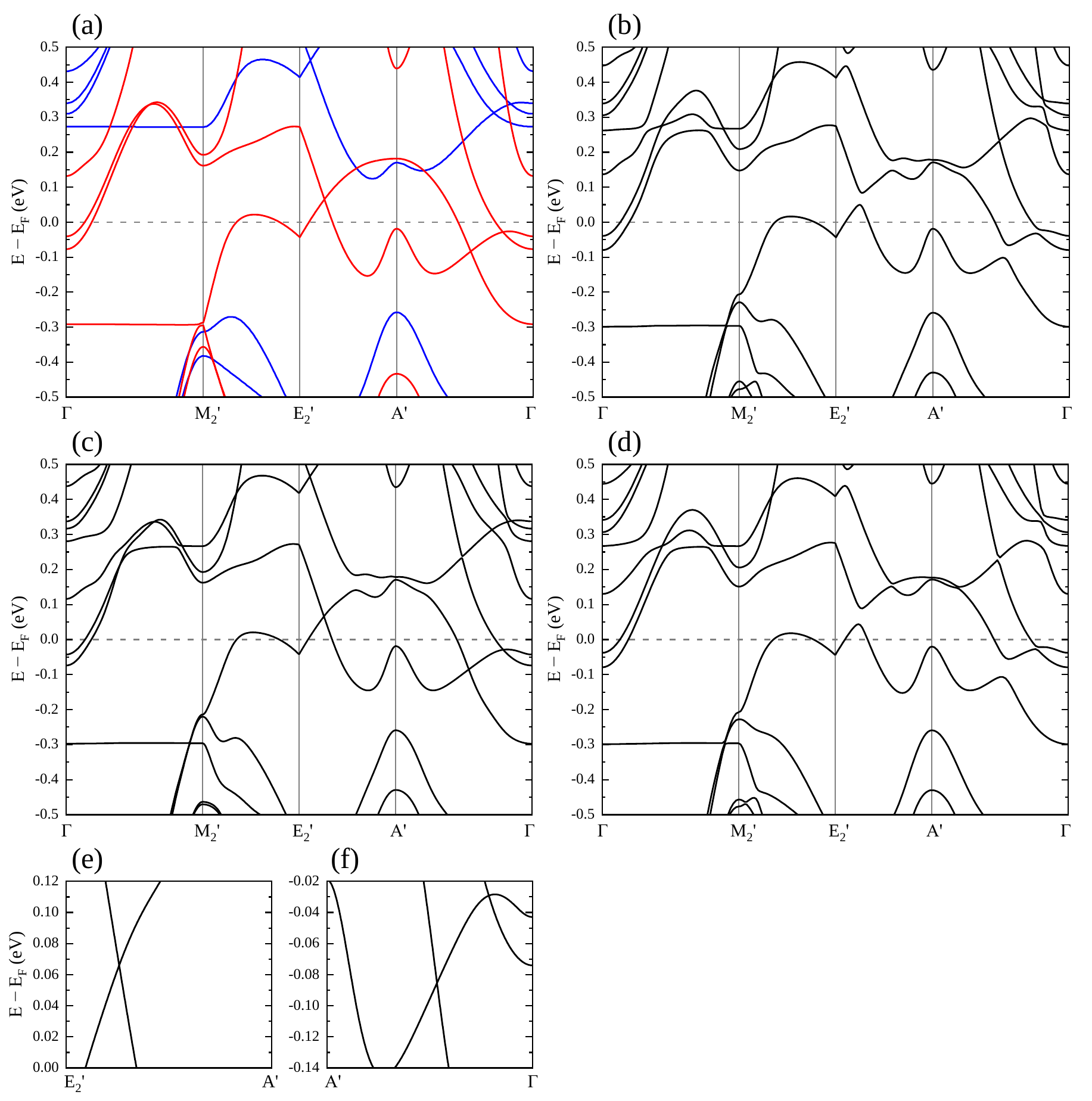}
\caption{Electronic band structures in FM states. (a) without SOC (red and blue lines are for majority spin bands and minority spin bands, respectively), (b)  FM$_{a}$ with SOC (Co spins along the $a_{m}$ axis), (c) FM$_{b}$with SOC (Co spins along the $b_{m}$ axis), and (d)  FM$_{c}$ with SOC (Co spins along the $c_{m}$  axis). (e) and (f) are magnified views of panel (c) near the nodal ring.  }
\label{fig:SX1}
\end{figure*}

\begin{figure*}
\centering
\includegraphics[width=1\linewidth]{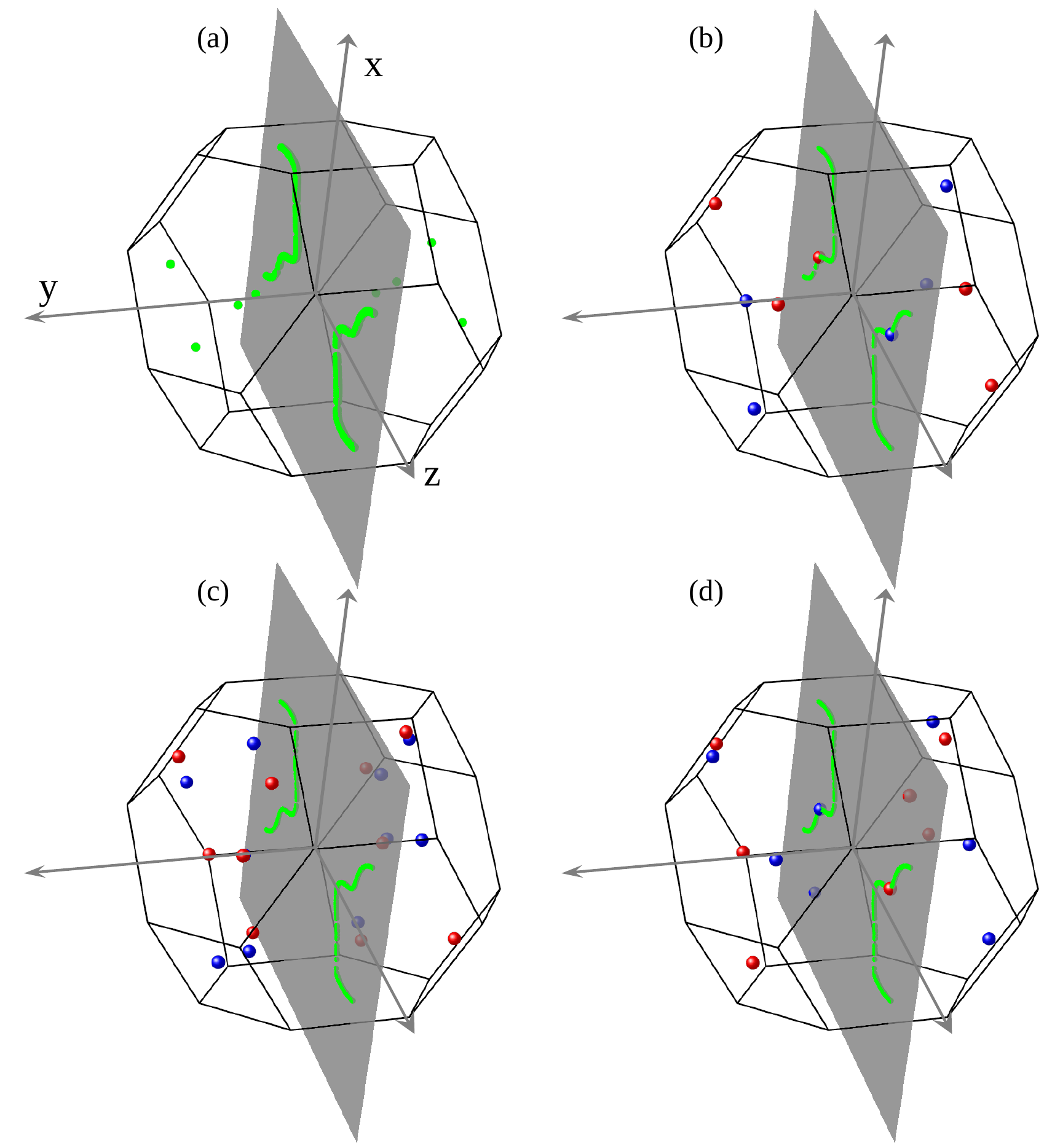}
\caption{Nodal lines, Dirac points, and Weyl points in FM states. (a) Results without SOC. Green lines in the mirror plane (gray plane) form a nodal ring. Green dots away from the mirror plane are Dirac points that are not protected by the mirror symmetry. (b) results for FM$_{a}$ with SOC, (c) FM$_{b}$, and (d), FM$_{c}$. In panels (b) and (d), green lines are the nodal lines appearing without SOC. Red (blue) spheres are Weyl points with positive (negative) chirality. In these cases, there is one pair of Weyl points in the mirror plane or on the nodal ring without SOC. Away from the mirror plane, there are four pairs of Weyl points in FM$_{a}$ and six pairs of Weyl points in  FM$_{c}$. In  FM$_{b}$, nodal ring is protected by the mirror symmetry as indicated by green lines in panel (c), with additional ten pairs of Weyl points away from the mirror plane. }
\label{fig:SX1}
\end{figure*}